\documentclass[usenatbib, longauth,letter]{aa} 

\usepackage{graphicx}
\usepackage{graphics}
\usepackage{txfonts}
\usepackage{epsfig}
\usepackage{natbib}
\usepackage{float}
\usepackage{hyperref}
\usepackage{dblfloatfix}
\usepackage[]{color}

\begin{document}

\title{Detection of the Schwarzschild precession in the orbit of the star S2 near the Galactic centre massive black hole}

\titlerunning{Detection of Schwarzschild Precession}
\subtitle{}

\author{GRAVITY Collaboration\thanks{GRAVITY is developed
    in a collaboration by the Max Planck Institute for
    extraterrestrial Physics, LESIA of Observatoire de Paris/Universit\'e PSL/CNRS/Sorbonne Universit\'e/Universit\'e de Paris and IPAG of Universit\'e Grenoble Alpes /
    CNRS, the Max Planck Institute for Astronomy, the University of
    Cologne, the CENTRA - Centro de Astrofisica e Gravita\c c\~ao, and
    the European Southern Observatory. $\,\,\,\,\,\,$ Corresponding authors: R.~Genzel (genzel@mpe.mpg.de), S.~Gillessen (ste@mpe.mpg.de), A.~Eckart (eckart@ph1.uni-koeln.de).
    }:
R.~Abuter\inst{8}
\and A.~Amorim\inst{6,13}
\and M.~Baub\"ock\inst{1}
\and J.P.~Berger\inst{5,8}
\and H.~Bonnet\inst{8}
\and W.~Brandner\inst{3}
\and V.~Cardoso\inst{13, 15}
\and Y.~Cl\'{e}net\inst{2}
\and P.T.~de~Zeeuw\inst{11,1}
\and J.~Dexter\inst{14,1}
\and A.~Eckart\inst{4,10}
\and F.~Eisenhauer\inst{1}
\and N.M.~F\"orster~Schreiber\inst{1} 
\and P.~Garcia\inst{7,13}
\and F.~Gao\inst{1}
\and E.~Gendron\inst{2}
\and R.~Genzel\inst{1,12}
\and S.~Gillessen\inst{1}
\and M.~Habibi\inst{1}
\and X.~Haubois\inst{9}
\and T.~Henning\inst{3}
\and S.~Hippler\inst{3}
\and M.~Horrobin\inst{4}
\and A.~Jim\'enez-Rosales\inst{1}
\and L.~Jochum\inst{9}
\and L.~Jocou\inst{5}
\and A.~Kaufer\inst{9}
\and P.~Kervella\inst{2}
\and S.~Lacour\inst{2}
\and V.~Lapeyr\`ere\inst{2}
\and J.-B.~Le~Bouquin\inst{5}
\and P.~L\'ena\inst{2}
\and M.~Nowak\inst{17,2}
\and T.~Ott\inst{1}
\and T.~Paumard\inst{2}
\and K.~Perraut\inst{5}
\and G.~Perrin\inst{2}
\and O.~Pfuhl\inst{8,1}
\and G.~Rodr\'iguez-Coira\inst{2}
\and J.~Shangguan\inst{1}
\and S.~Scheithauer\inst{3}
\and J.~Stadler\inst{1}
\and O.~Straub\inst{1}
\and C.~Straubmeier\inst{4}
\and E.~Sturm\inst{1}
\and L.J.~Tacconi\inst{1}
\and F.~Vincent\inst{2}
\and S.~von~Fellenberg\inst{1}
\and I.~Waisberg\inst{16,1}
\and F.~Widmann\inst{1}
\and E.~Wieprecht\inst{1}
\and E.~Wiezorrek\inst{1} 
\and J.~Woillez\inst{8}
\and S.~Yazici\inst{1,4}
\and G.~Zins\inst{9}
}

\institute{
Max Planck Institute for extraterrestrial Physics,
Giessenbachstra{\ss}e~1, 85748 Garching, Germany
\and LESIA, Observatoire de Paris, Universit\'e PSL, CNRS, Sorbonne Universit\'e, Universit\'e de Paris, 5 place Jules Janssen, 92195 Meudon, France
\and Max Planck Institute for Astronomy, K\"onigstuhl 17, 
69117 Heidelberg, Germany
\and $1^{\rm st}$ Institute of Physics, University of Cologne,
Z\"ulpicher Stra{\ss}e 77, 50937 Cologne, Germany
\and Univ. Grenoble Alpes, CNRS, IPAG, 38000 Grenoble, France
\and Universidade de Lisboa - Faculdade de Ci\^encias, Campo Grande,
1749-016 Lisboa, Portugal 
\and Faculdade de Engenharia, Universidade do Porto, rua Dr. Roberto
Frias, 4200-465 Porto, Portugal 
\and European Southern Observatory, Karl-Schwarzschild-Stra{\ss}e 2, 85748
Garching, Germany
\and European Southern Observatory, Casilla 19001, Santiago 19, Chile
\and Max Planck Institute for Radio Astronomy, Auf dem H\"ugel 69, 53121
Bonn, Germany
\and Sterrewacht Leiden, Leiden University, Postbus 9513, 2300 RA
Leiden, The Netherlands
\and Departments of Physics and Astronomy, Le Conte Hall, University
of California, Berkeley, CA 94720, USA
\and CENTRA - Centro de Astrof\'{\i}sica e
Gravita\c c\~ao, IST, Universidade de Lisboa, 1049-001 Lisboa,
Portugal
\and Department of Astrophysical \& Planetary Sciences, JILA, Duane Physics Bldg., 2000 Colorado Ave, University of Colorado, Boulder, CO 80309, USA
\and CERN, 1 Espl. des Particules, Gen\`eve 23, CH-1211, Switzerland
\and Department of Particle Physics \& Astrophysics, Weizmann Institute of Science, Rehovot 76100, Israel
\and Institute of Astronomy, Madingley Road, Cambridge CB3 0HA, UK
}

\date{Draft version \today}

\abstract{The star S2 orbiting the compact radio source Sgr~A* is a precision probe of the gravitational field around the closest massive black hole (candidate). Over the last 2.7 decades we have monitored the star's radial velocity and motion on the sky, mainly with the SINFONI and NACO adaptive optics (AO) instruments on the ESO VLT, and since 2017, with the four-telescope interferometric beam combiner instrument GRAVITY. In this paper we report the first detection of the General Relativity (GR) Schwarzschild Precession (SP) in S2's orbit. Owing to its highly elliptical orbit ($e=0.88$), S2's SP is mainly a kink between the pre-and post-pericentre directions of motion $\approx \pm 1\,$year around pericentre passage, relative to the corresponding Kepler orbit. The superb 2017-2019 astrometry of GRAVITY defines the pericentre passage and outgoing direction.  The incoming direction is anchored by 118 NACO-AO measurements of S2's position in the infrared reference frame, with an additional 75 direct measurements of the S2-Sgr~A* separation during bright states (`flares') of Sgr~A*. Our 14-parameter model fits for the distance, central mass, the position and motion of the reference frame of the AO astrometry relative to the mass, the six parameters of the orbit, as well as a dimensionless parameter $f_\mathrm{SP}$ for the SP ($f_\mathrm{SP}=0$  for Newton and 1 for GR). From data up to the end of 2019 we robustly detect the SP of S2, $\delta \phi \approx 12'$ per orbital period. From posterior fitting and MCMC Bayesian analysis with different weighting schemes and bootstrapping we find $f_\mathrm{SP}=1.10 \pm 0.19$. The S2 data are fully consistent with GR. Any extended mass inside S2's orbit cannot exceed $\approx0.1$\% of the central mass. Any compact third mass inside the central arcsecond must be less than about $1000\,M_\odot$.}

\keywords{black hole physics -- Galaxy: nucleus  --  gravitation -- relativistic processes}

\maketitle

\section{Introduction}
\label{sec:intro}
\subsection{Testing GR and the massive black hole paradigm}
The theory of General Relativity (GR) continues to pass all experimental tests with flying colours \citep{1916AnP...354..769E, 2014LRR....17....4W}. High-precision laboratory and Solar System experiments, and observations of solar-mass pulsars in binary systems \citep{2006Sci...314...97K, 2016IJMPD..2530029K} have confirmed GR in the low-curvature regime. Gravitational waves from several stellar mass, black hole (sBH) candidate in-spirals with LIGO \citep{2016PhRvL.116f1102A} have tested the strong-curvature limit. 

General Relativity predicts black holes, that is, space-time solutions with a non-spinning or spinning central singularity cloaked by a communication barrier, an event horizon (cf. \citealt{1916SPAW.......189S, 1965qssg.conf...99K}). The LIGO measurements currently provide the best evidence that the compact in-spiralling binaries are indeed merging sBHs, but see \cite{2019LRR....22....4C}. 

Following the discovery of quasars \citep{1963Natur.197.1040S}, evidence has been growing that most massive galaxies harbour a central compact mass, perhaps in the form of a massive black hole (MBH: $10^6 - 10^{10} M_\odot$, \citealt{1971MNRAS.152..461L, 2013ARA&A..51..511K, 2013ApJ...764..184M}).  Are these compact mass concentrations truly MBHs, as predicted by GR? Evidence in favour comes from relativistically broadened, redshifted iron K$\alpha$ line emission in nearby Seyfert galaxies \citep{1995Natur.375..659T, 2000PASP..112.1145F}, from stellar or gas motions very close to them (e.g. \citealt{1999JApA...20..165M}), and high resolution millimetre imaging \citep{2019ApJ...875L...1E}.
 
The nearest MBH candidate is at the centre of the Milky Way ($R_0\approx 8\,$kpc, $M_\bullet \approx 4\times10^6 M_\odot$, \citealt{2010RvMP...82.3121G, 2008ApJ...689.1044G}). It is coincident with a very compact and variable X-ray, infrared, and radio source, Sgr~A*, which in turn is surrounded by a very dense cluster of orbiting young and old stars. Radio and infrared observations have provided detailed information on the distribution, kinematics, and physical properties of this nuclear star cluster and hot, warm, and cold interstellar gas interspersed in it (cf. \citealt{2010RvMP...82.3121G, 2012RAA....12..995M, 2013CQGra..30x4003F}). Groups in Europe at the ESO NTT \& VLT and in the USA at the Keck telescopes have carried out high-resolution imaging and spectroscopy of the nuclear star cluster over the past two and a half decades. They determined increasingly precise motions for more than $10^4$ stars, and orbits for $\approx 50$ \citep{2002Natur.419..694S, 2003ApJ...586L.127G, 2008ApJ...689.1044G, 2005ApJ...628..246E, 2009ApJ...692.1075G, 2009A&A...502...91S, 2012Sci...338...84M, 2016ApJ...830...17B, 2016ApJ...821...44F,  2017ApJ...837...30G}. These orbits, in particular the highly eccentric orbit of the $m_K \approx 14$ star S2 (or `S02' in the UCLA nomenclature), have demonstrated that the gravitational potential is dominated by a compact source of $4.25 \times 10^6 M_\odot$, concentrated within the pericentre distance of S2. S2 appears to be a slowly rotating, single, main-sequence B-star of age $\approx 6\,$Myr \citep{2008ApJ...672L.119M, 2017ApJ...847..120H, 2017A&A...602A..94G, 2018ApJ...854...12C}. 

The location of the radio source Sgr~A* coincides with that of the mass centroid to much better than $1\,$mas \citep{2015MNRAS.453.3234P, 2019ApJ...873...65S}. Millimetre Very Long Baseline Interferometry \citep{2000ApJ...528L..13F, 2008Natur.455...78D, 2017ApJ...850..172J, 2019ApJ...871...30I} shows that Sgr~A* has a $1.3\,$mm half-light radius smaller than $18\,\mu$as, or 1.8 times the Schwarzschild radius ($R_S$) of a $4.25 \times 10^6 M_\odot$ MBH. Sgr~A* shows no detectable intrinsic motion within the international celestial reference frame ICRF. This supports the interpretation that the compact radio source is coincident with the mass \citep{2004ApJ...616..872R, Reid:2009p2548, 2020arXiv200104386R}. 
The Galactic Centre (GC) currently provides the best `laboratory' for testing GR near MBHs and ultimately for testing the MBH paradigm \citep{2005PhR...419...65A, 2017ARA&A..55...17A, 2010RvMP...82.3121G, 2016ApJ...818..121P}.

\subsection{Detection of GR effects in the orbits of stars around Sgr~A*: Gravitational redshift }

Following the observations of the pericentre passage of S2 in 2002.33 \citep{2002Natur.419..694S, 2003ApJ...586L.127G}
it became clear that the first-order ($O(\beta^2),\,\beta=v/c)$ GR-effects of the orbit may be in reach of precision observations. These are the gravitational redshift (RS) $\mathrm{PPN1}_\mathrm{RS}(\lambda)$, and the Schwarzschild precession (SP) $\mathrm{PPN1}_\mathrm{SP}(x,y)$, see \cite{2001A&A...374...95R}, \cite{2006ApJ...639L..21Z}, \cite{2010ApJ...720.1303A}, \cite{2014MNRAS.444.3780A}, \cite{2017A&A...608A..60G}, and \cite{2017ApJ...845...22P}. For this purpose, a significant (factor $4-10$) improvement in astrometry compared to what was possible in 2010 was needed. We achieved this goal with the development of GRAVITY, a cryogenic, interferometric beam combiner of all four UTs of the ESO VLT, along with adaptive optics (AO) systems for all four UTs, and a laser metrology system \citep{2017A&A...602A..94G}.
 
On May 19, 2018 (2018.38), S2 passed pericentre at $120\,$AU ($\approx 1400\,R_S$) with an orbital speed of $7700\,$km/s ($\beta=0.026$). From monitoring the star's radial velocity and motion on the sky from data taken prior to and up to two months after pericentre, \cite{2018A&A...615L..15G} were able to detect the first post-Newtonian effect of GR, the gravitational redshift, along with the transverse Doppler effect of special relativity (SRT, \citealt{1973grav.book.....M}). \cite{2019A&A...625L..10G} improved the statistical robustness of the detection of the RS to $f_\mathrm{RS}=1.04\pm0.05$, where the dimensionless parameter $f_\mathrm{RS}$ is 0 for Newtonian orbits and 1 for GR-orbits. \cite{2019Sci...365..664D} confirmed these findings from a second, independent data set mainly from the Keck telescope, $f_\mathrm{RS}= 0.88 \pm 0.17$.
 
The combined $\mathrm{PPN1}_\mathrm{RS}(\lambda)$ gravitational redshift and transverse Doppler effect are detected as a $200\,$km/s residual centred on the pericentre time, relative to the $f_\mathrm{RS}=0$ orbit (with the same other parameters describing the star's orbit and the gravitational potential). While the RS occurs solely in wavelength-space, the superior astrometric capabilities of GRAVITY serve to set much tighter constraints on the orbital geometry, mass and distance, thus decreasing the uncertainty of $f_\mathrm{RS}$  more than three times relative to data sets constructed from single-telescope, AO imaging and spectroscopy.
 
In the following we report the first measurement of the next relativistic effect in the orbit of S2, namely the in-plane, prograde precession of its pericentre angle, the Schwarzschild precession \citep{1973grav.book.....M}.

\section{Observations}
\label{sec:observations}

\begin{figure*}[t!]
\centering
\begin{minipage}{.655\textwidth}
\includegraphics[width=\linewidth]{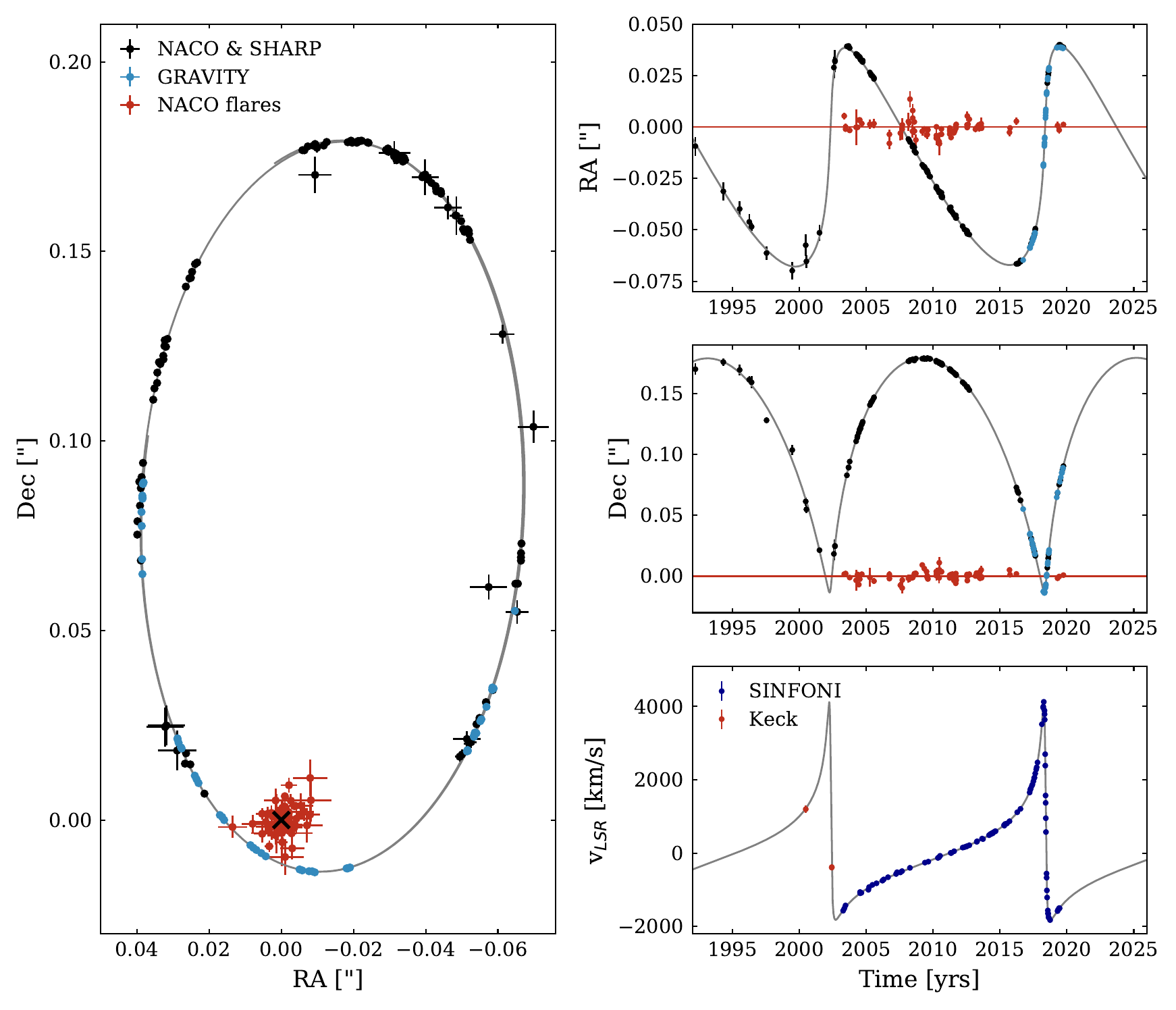}
\end{minipage} \quad
\begin{minipage}{.3\textwidth}
\caption{Summary of the observational results of monitoring the S2-Sgr~A* orbit from 1992 to the end of 2019. Left: SHARP, NACO (black points), and GRAVITY (blue points) astrometric positions of the star S2, along with the best-fitting GR orbit (grey line). The orbit does not close as a result of the SP. The mass centre is at (0,0), marked by the cross. All NACO and SHARP points were corrected for a zero-point offset and drift in RA and Dec. The red data points mark the positions of the infrared emission from Sgr~A* during bright states, where the separation of S2 and Sgr~A* can be directly inferred from differential imaging. Right: RA (top) and Dec (middle) offset of S2 (black and blue) and of the infrared emission from Sgr~A* (red) relative to the position of Sgr~A* (assumed to be identical with the mass centre). Grey is the best-fitting GR-orbit including the R{\o}mer effect (finite speed of light), SRT, and GR to PPN1. We assumed $f_\mathrm{RS}=1$ and fitted for $f_\mathrm{SP}$. Bottom right: Same for the line-of-sight velocity of the star.}
\label{fig:fig1} 
\end{minipage}
\end{figure*}

Following on from \cite{2018A&A...615L..15G, 2019A&A...625L..10G}, we expand in this paper our analysis of the positions and K-band spectra of the star S2 by another year, to fall of 2019. This yielded 5 additional NACO points, 6 SINFONI points and, especially, 11 crucial GRAVITY points. We now have 

\begin{itemize}
\item 118 measurements with the VLT AO-assisted infrared camera NACO \citep{1998SPIE.3354..606L, 1998SPIE.3353..508R} between 2002 and 2019.7 of the position of S2 in the K or H bands, relative to the `Galactic Centre infrared reference system' (\citealt{2015MNRAS.453.3234P}, rms uncertainty $\approx 400\,\mu$as). This means that between the 2002.33 pericentre passage until 2019.7 we have 7 to 16 NACO positional measurements per year. Between 1992 and 2002, we also used the speckle camera SHARP at the NTT \citep{1993Ap&SS.205....1H}, but the astrometry of the speckle data on a $3.5\,$m telescope is an order of magnitude worse than the AO imagery on the $8\,$m VLT (rms uncertainty $\approx 3.8\,$mas); 
\item 75 NACO measurements between 2003.3 and 2019.7 of the direct S2-Sgr~A* separation during bright states of Sgr~A* (typical rms uncertainty $1.7\,$mas);
\item 54 GRAVITY measurements between 2016.7 and 2019.7 of the S2-Sgr~A* separation (rms uncertainty $\approx 65\,\mu$as). During the pericentre-passage year 2018, the sampling was especially dense with 25  measurements;
\item 92 spectroscopic measurements of the $2.167\,\mu$m HI (Br$\gamma$) and the $2.11\,\mu$m HeI lines between 2003.3 and 2019.45 with the AO-assisted integral field spectrometer SINFONI at the VLT \citep{2003SPIE.4841.1548E, 2003SPIE.4839..329B}, with an uncertainty of $\approx12\,$km/s \citep{2019A&A...625L..10G}. This means that we typically have 3 to 6 spectroscopic measurements per year, and more than 20 in 2018. We also added 2 more NACO AO slit spectroscopic measurements from 2003, and 3 more Keck-NIRC2 AO spectroscopic measurements between 2000 and 2002 \citep{2019Sci...365..664D}.
\end{itemize}

The SHARP and NACO data deliver relative positions between stars in the nuclear star cluster, which are then registered in the radio frame of the GC \citep{Reid:2009p2548, 2020arXiv200104386R} by multi-epoch observations of nine stars in common between the infrared and radio bands. Another important step is the correction for spatially variable image distortions in the NACO imager, which are obtained from observations of an HST-calibrated globular cluster \citep{2015MNRAS.453.3234P}. The radio calibrations still allow for a zero-point offset and a drift of the radio-reference frame centred on Sgr~A* (strictly speaking, on the mass-centroid) with respect to the infrared reference frame, which we solve for empirically in our orbit fitting. For this purpose, we use the \cite{2015MNRAS.453.3234P} radio-to-infrared reference frame results as a prior ($x_0 = -0.2 \pm 0.2\,$mas, $y_0 = 0.1 \pm 0.2\,$mas, $vx_0 = 0.05 \pm 0.1\,$mas/yr, $vy_0 = 0.06 \pm 0.1\,$mas/yr). These reference frame parameters $(x_0,\,y_0,\,vx_0$, and $vy_0)$ are now the limiting factor in the precision of the detection of the SP of S2.

The situation is different for GRAVITY. Here we detect and stabilise the interferometric fringes on the star IRS16C located $\approx1''$ NE of Sgr~A*, and observe S2 or Sgr~A* within the second phase-referenced fibre (see \citealt{2017A&A...602A..94G}), such that the positional difference between S2 and Sgr~A* can be determined to $<100\,\mu$as levels (see Appendix~\ref{sec:A1}). To obtain this accuracy, the measurements of S2 and Sgr~A* are made within a short time interval and linked together interferometrically (Appendix~\ref{sec:A2}). Between the end of 2017 and throughout 2018, S2 and Sgr~A* are simultaneously detected in a single fibre-beam positioning as two unresolved sources in $>95$\% of our individual integrations (5 min each), such that the S2-Sgr~A* distance is even more directly obtained in each of these measurements (Appendix~\ref{sec:A3}). The development over time of the astrometric and spectroscopic measurement uncertainties are summarised in Figure~\ref{fig:figA3}. For more details on the data analysis of all three instruments we refer to \cite{2017A&A...602A..94G, 2018A&A...615L..15G, 2018A&A...618L..10G, 2019A&A...625L..10G} and Appendix~\ref{sec:A}.

\section{Results}
\label{sec:results}

\subsection{Schwarzschild precession in the S2 orbit}

Figure~\ref{fig:fig1} shows the combined single-telescope and interferometric astrometry of the 1992-2019 sky-projected orbital motion of S2 and the line-of sight velocity of the star. 
The almost 100-fold improvement of statistical astrometric measurement precision in the past 27 years is one key for detecting the SP in the S2 orbit. As discussed in Section~\ref{sec:observations}, the accurate definition of the reference frame for the NACO data is the second key. The robustness of the detection of the SP strongly correlates with the precision of knowing $(x_0,\, y_0,\, vx_0$, and $vy_0)$, as this sets the angle of the orbit at the last apocentre (2010.35). Using the priors from \cite{2015MNRAS.453.3234P}, we fitted these four reference frame parameters in our posterior fitting, but we found the additional constraints obtained from Sgr~A*-S2 flare offsets in NACO to be very helpful. To this end, we included in the calculation of $\chi^2$ the constraint that the flare positions are tracing the mass centre.

Confusion of S2 with nearby other sources is the final key issue (see also \citealt{2009ApJ...692.1075G,2017ApJ...837...30G,2018MNRAS.476.4372P,2019Sci...365..664D}). \cite{2003ApJ...586L.127G} and \cite{2002Natur.419..694S} already have noted that the NACO or NIRC2 AO astrometry at times was unreliable and biased over longer periods of time ($0.5-1.5\,$years). These systematic position excursions are mainly caused by confusion, that is, the positional pulling of the apparent sky position of S2 by a passing nearby background object. This issue is especially detrimental when the variable Sgr~A* emission source is within the diffraction limit of the telescope \citep{2003ApJ...586L.127G, 2008ApJ...689.1044G, 2018MNRAS.476.4372P, 2019Sci...365..664D}, making the 2002 and 2018 AO astrometry more uncertain or even unusable. Fortunately, GRAVITY removed any need for AO imagery during the 2018 pericentre passage, therefore we excised most of the 2002 and 2018 NACO astrometry from our data set. We identified further confusion events with fainter stars passing close to S2 on a number of occasions (e.g. 1998, 2006, and 2013/2014) and removed these questionable data points.
 
 \begin{figure*}[ht]
\centering
\begin{minipage}{.72\textwidth}
\includegraphics[width=\linewidth]{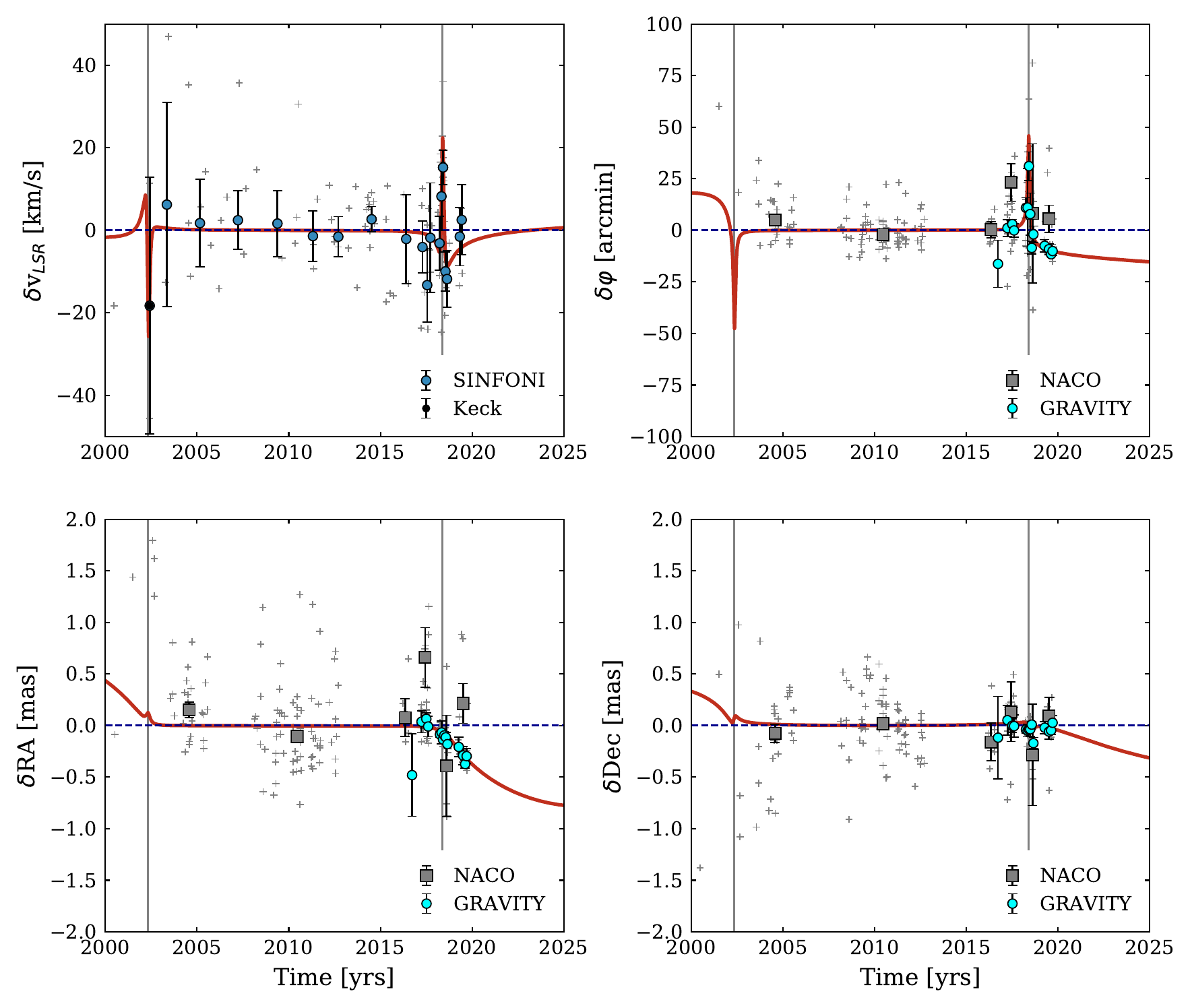}
\end{minipage} \quad
\begin{minipage}{.25\textwidth}
\caption{Posterior analysis of all data by fitting for $f_\mathrm{SP}$ simultaneously with all other parameters. As in Figure~\ref{fig:figB2}, the bottom panels show the residuals in RA (left) and Dec (right) between the data and the best-fitting GR (thick red curve, $f_\mathrm{SP} =1.1$), relative to the same orbit for $f_\mathrm{SP}=0$  (Newton, plus R{\o}mer effect, plus SRT, plus RS). Grey crosses denote individual NACO or SINFONI data, cyan filled black circles show averaged GRAVITY data, and grey rectangles denote averages of the NACO data. The top right panel shows the same for $\delta \varphi$, and the top left panel for $\delta vz$. Blue filled black circles are averages of the SINFONI residuals, with all residuals shown as grey crosses. The best fit (red curve) including the flare data (Figure~\ref{fig:fig1}) has $f_\mathrm{SP}=1.1$, with a $1\sigma$ uncertainty of $0.19$. The overall reduced $\chi^2_r$  of this fit is $1.5$.}
\label{fig:fig2}
\end{minipage}
\end{figure*}

At pericentre $R_\mathrm{peri}$, S2 moves with a total space velocity of $\approx 7700\,$km/s, or $\beta = v/c = 2.56 \times 10^{-2}$. The SP of the orbit is a first-order ($\beta^{2N}, N=1$) effect in the parametrised post-Newtonian (PPN, cf. \citealt{1972ApJ...177..757W}) expansion, PPN(1)$\,\approx \beta^2 \approx  R_S / R_\mathrm{peri} \approx 6.6 \times 10^{-4}$. We used the post-Newtonian expansion of \cite{2008ApJ...674L..25W} and added a factor $f_\mathrm{SP}$ in the equation of motion in front of the Schwarzschild-related terms (see Appendix~\ref{sec:PPN}). This corresponds to (e.g. \citealt{1973grav.book.....M})
\begin{equation}
\Delta \phi_\mathrm{per\,orbit}=\mathrm{PPN1}_\mathrm{SP} = f_\mathrm{SP} \frac{3 \pi R_S}{a(1-e^2)} \, \stackrel{\mathrm{for\,S2}}{=}  \, f_\mathrm{SP} \times 12.1'\,\,.
\end{equation}
Here $a$ is the semi-major axis and $e$ is the eccentricity of the orbit. The quantity $f_\mathrm{SP}$ can then be used as a fitting parameter, similar to our approach for the RS \citep{2018A&A...618L..10G, 2019A&A...625L..10G}. 
Appendix~\ref{theo} explains the effects that the SP should have on the measured parameters of the S2 orbit.
 
\subsection{Posterior analysis}

The six parameters describing the Kepler orbit $(a,\, e,\, i,\, \omega,\, \Omega$, and $ t_0)$, the distance, and the central mass, and the five coordinates describing the position on the sky and the three-dimensional velocity of the reference frame (relative to the AO spectroscopic or imaging frames) all have uncertainties. In particular, distance and mass are uncertain and correlated. Following \cite{2018A&A...615L..15G, 2019A&A...625L..10G}, we determined the best-fit value of the parameter  $f_\mathrm{SP}$ a posteriori, including all data and fitting for the optimum values of all parameters with the Levenberg-Marquardt $\chi^2$-minimisation algorithm \citep{1944L, 1963M}, including prior constraints. It is essential to realise that the inferred measurement uncertainties are affected and partially dominated by systematic effects, especially when the evidence from three or more very different measurement techniques is combined. 

\begin{figure*}[ht]
\centering
\begin{minipage}{.72\textwidth}
\includegraphics[width=\linewidth]{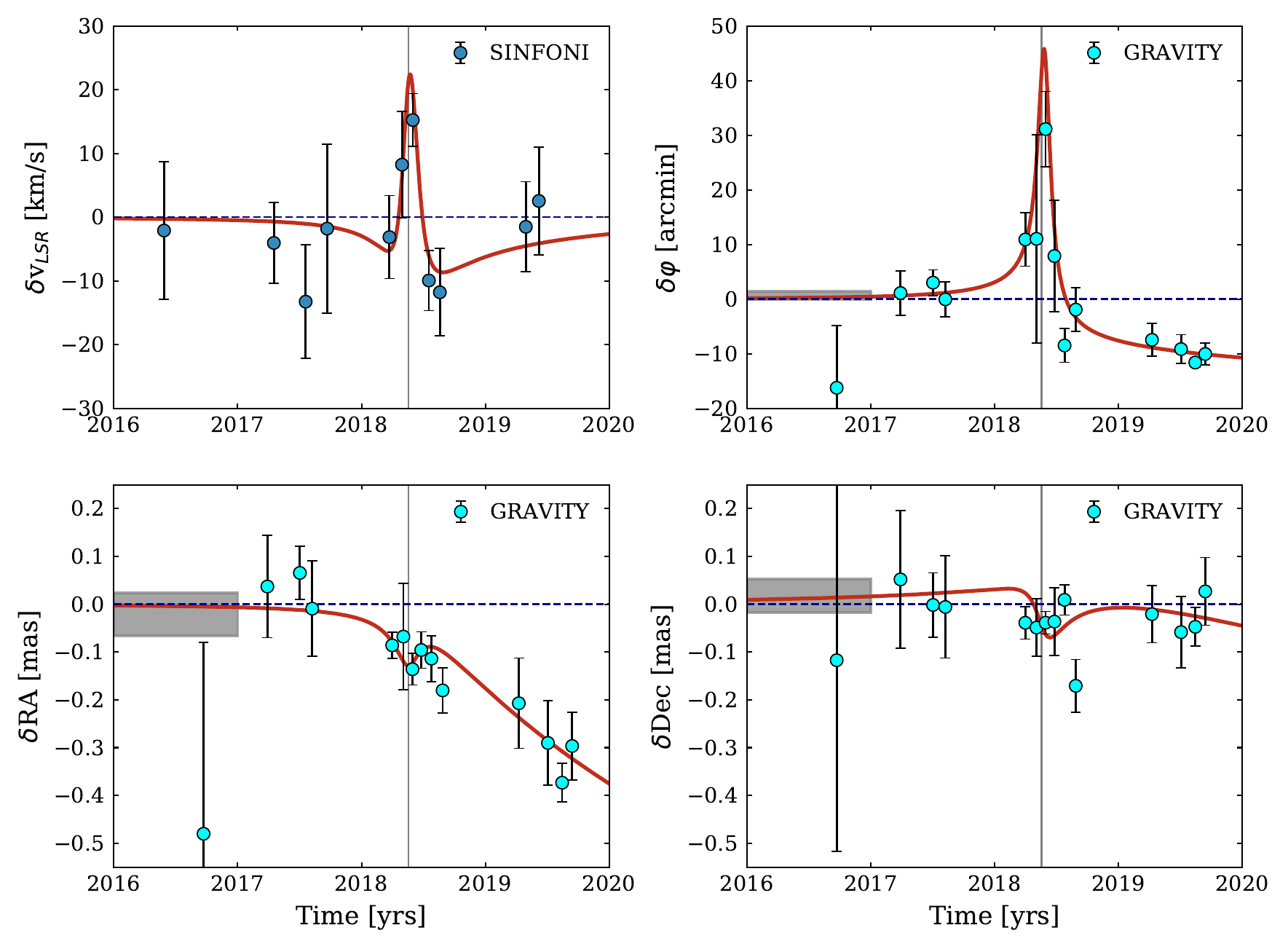}
\end{minipage} \quad
\begin{minipage}{.25\textwidth}
\caption{As Figure~\ref{fig:fig2}, but now zooming into the critical 2018 pericentre residuals (best-fit $f_\mathrm{SP}=1.1$ minus  $f_\mathrm{SP}=0$, with all other parameters fixed). In the bottom two and the top right panels, blue filled, open black circles (with $1\sigma$ uncertainties) are averages of GRAVITY data. The grey bar on the left denotes the constraints obtained from NACO imagery of S2 and Sgr~A* flares on the location of the apocentre value (2010.35). Averages (and $1\sigma$ uncertainties) of the radial velocity residuals from SINFONI are open black, blue filled circles (top left). The residuals of the best-fitting $f_\mathrm{SP}=1.1$ minus  $f_\mathrm{SP}=0$ curves are plotted in red.}
\label{fig:fig3}
\end{minipage} 
\end{figure*}

Figures~\ref{fig:fig2} and~\ref{fig:fig3} show the fit results when we simultaneously fitted the data and the flare positions. 
As priors we used the  \cite{2015MNRAS.453.3234P} reference frame results (see Section~\ref{sec:observations}). All data prior to the 2018.3 pericentre passage are fit by $f_\mathrm{SP}\approx0$ (Newton, plus R{\o}mer effect, plus SRT, plus RS). The residuals in this period are consistent with 0 (bottom panels of Figure~\ref{fig:fig2}). The GRAVITY data between 2017 and 2019.7  clearly show that the post-pericentre orbit exhibits a sudden kink, mainly in RA. The data are compatible with a pure in-plane precession. This is shown in the upper right panels of Figures~\ref{fig:fig2} and~\ref{fig:fig3}, where we have computed the residuals in the projected angle of the SP-fitted orbit on the sky $\delta \varphi(t)$, relative to the $f_\mathrm{SP}=0$ orbit. This is exactly as expected from an $f_\mathrm{SP}\approx1$ GR orbit (Figure~\ref{fig:figB2}). The more subtle swings in $\delta$RA, $\delta$Dec, $\delta vz$, and $\delta \varphi$ predicted by GR (Figure~\ref{fig:figB2}) are detected as well (see Appendix~\ref{theo} for a more detailed discussion). 

Table~\ref{tab:t1} lists the best-fit parameters and their $1\sigma$ uncertainties. Depending on the weighting of different data sets and the choice of priors, we find that the best-fitting $f_\mathrm{SP}$ value varies between 0.9 and 1.2, with a fiducial value of  $f_\mathrm{SP}=1.1$. The formal statistical fit uncertainty of this parameter does not depend much on the selection of astrometric and spectroscopic data of S2. The value of its rms uncertainty $\Delta f_\mathrm{SP}$ does depend on the methodology of error treatment. The distribution of the NACO flare position residuals shows significant non-Gaussian outliers. There are $\approx$ six (of 75 data points) $>4\sigma$ outliers above the rms of $\approx 1.7\,$mas. If the $\chi^2$ distribution and the weighting of these points are treated as if they had a normal distribution, the reduced  $\chi^2_r$ of our overall fits is driven up to $\approx1.65$, for a total  $\chi^2$ of 995. In this case $\Delta f_\mathrm{SP} = 0.204$. These outliers can be down-weighted by replacing the penalty function $p(r) = r^2$ in the calculation of $\chi^2 = \Sigma p( (\mathrm{data}-\mathrm{model}) / \mathrm{error})$ with $p(r,s) =  r^2 \cdot s^2 / (r^2 + s^2), s = 10$. This introduces a soft cut-off around $10\sigma$ in how much a data point can maximally contribute to the $\chi^2$. With this scheme, $\chi^2_r$ of the overall fit drops to 1.50, and  $\Delta f_\mathrm{SP} = 0.194$.

We also fitted the data by solving simultaneously for  $f_\mathrm{SP}$  and  $f_\mathrm{RS}$, without fixing the RS term to 1. In this case we find  $f_\mathrm{SP} = 0.99 \pm 0.24$ and  $f_\mathrm{RS}=0.965 \pm 0.042$ (with the outlier damper on), again fully consistent with GR.
  
An alternative approach is to place the reference frame constraints we obtained from the flare positions into a combined prior with the contsraints from \cite{2015MNRAS.453.3234P}. In this case the prior for the location of Sgr~A* in the NACO infrared frame is $x_0 = -0.42 \pm 0.15\,$mas, $y_0 = 0.30 \pm 0.15\,$mas, $vx_0 = -0.02 \pm 0.05\,$mas/yr, and $vy_0 = 0.015 \pm 0.05\,$mas/yr. When we use this prior to fit only the S2 data, we obtain $f_\mathrm{SP}  = 0.92 \pm 0.22$ and $\chi^2_r= 0.88\,\, (\chi^2 = 398)$.

Fitting the orbit with $f_\mathrm{SP} =0$ fixed yields $\chi^2 =  932.3$, compared to 906.4 with $f_\mathrm{SP}=1$ fixed. The corresponding difference in Bayesian information criterion \citep{CH2008} $\Delta \mathrm{BIC} = 25.9$ yields very strong evidence that the GR model describes the data better than the best-fitting Kepler (with SRT, RS, and R{\o}mer delay included) orbit.
 
\cite{2018A&A...618L..10G} showed that the near-infrared emission of Sgr~A* during bright flares exhibits clock-wise loop motions of excursions $50-100\,\mu$as. The typical flare duration and the orbital timescale are $\approx 1\,$hour.
 A stationary offset between the infrared emission and the mass centroid of that size would induce a change of up to $\pm0.2$ in $f_\mathrm{SP}$, comparable to the overall uncertainty in $f_\mathrm{SP}$. 
During a typical time of several hours making up a GRAVITY data point in this work these fluctuations should average out to less than $10\,\mu$as such that the additional error on $f_\mathrm{SP}$ is well below the statistical error. 
  
Next we carried out a Markov-Chain Monte Carlo (MCMC) analysis. Using $200,000$ realisations we found that the distribution of $f_\mathrm{SP}$  is well described by a Gaussian centred on $f_\mathrm{SP}  = 1.11 \pm 0.21$ (Figure~\ref{fig:figE1}, and see Appendix~\ref{details} for more details). The largest relative uncertainty in the determination of the Schwarzschild term originates in the degeneracy of $f_\mathrm{SP}$ with the pericentre time (see Appendix~\ref{theo}) and with the zero-point $x_0$ of the long-term reference frame (mass vs. NACO imaging coordinates). This is expected because the precession is largest in the EW direction. 
  
Furthermore, we compared our first-order post-Newtonian code with fully relativistic GR orbits using the GYOTO ray-tracing code\footnote{Freely available at \hyperref[]{http://gyoto.obspm.fr}} \citep{2011CQGra..28v5011V, 2017A&A...608A..60G}. As expected, the deviations are small. The largest differences over the full data range  are $\Delta \mathrm{RA} = 62\,\mu$as, $\Delta \mathrm{Dec} = 41\,\mu$as and $\Delta vz = 11.4\,$km/s, occurring for a short time around pericentre. Moreover, the Bayesian comparison between the best-fitting full-GR and Kepler (with SRT, RS, and R{\o}mer delay included) orbits strongly prefers the GR model.

Finally, we also included the data from \cite{2019Sci...365..664D} (excepting the 2018 astrometry) using the scheme in \cite{2009ApJ...707L.114G} allowing for an additional offset in position and velocity for the Keck reference system. The 18-parameter fit yields a consistent result, but no further improvement.
   
\section{Conclusions}
\label{sec:conclusion}

We have presented the first direct detection of the in-plane Schwarzschild precession around Sgr~A* in the GC. Our results are fully consistent with GR. We detect the precession of S2 robustly at the 5 to $6\sigma$ level in posterior fitting and MCMC analysis. Our result is independent of the fit methodology, data selection, weighting, and error assignments of individual data points. The significance of our result depends mostly on how accurately we can constrain $(x_0,\, y_0,\, vx_0$, and $vy_0)$. The success rests crucially on the superior GRAVITY astrometry during and past pericentre passage on the one hand, and on 75 measurements of Sgr~A* flares from NACO AO data between 2003 and 2019 on the other. The flare data allow us to independently constrain the zero-point of the NACO reference frame. 

Additional masses in the GC would lead to Newtonian perturbations of the S2 orbit. An extended mass component (e.g. composed of stars or remnants, but also of other particles) would result in a retrograde precession. The presence of a second massive object would lead to short excursions in the smooth orbit figure. Our data place tight constraints on both, which we detail in Appendix~\ref{sec:discussion}, where we discuss several important astrophysical implications of our measurements.

We expect only modest further improvement of the significance of our result as our monitoring continues and S2 moves away from pericentre, because our result already now is limited by the precision with which we have measured the pre-pericentre orbit with AO data. 

\newpage
\begin{acknowledgements}

We are very grateful to our funding agencies (MPG, ERC, CNRS [PNCG, PNGRAM], DFG, BMBF, Paris Observatory [CS, PhyFOG], Observatoire des Sciences de l'Univers de
Grenoble, and the Funda\c c\~ao para a Ci\^encia e Tecnologia), to ESO and the ESO/Paranal staff, and to the many scientific and technical staff members in our institutions, who helped to make NACO, SINFONI, and GRAVITY a reality. S.G. acknowledges the support from ERC starting grant No. 306311. F.E. and O.P. acknowledge the support from ERC synergy grant No. 610058. A.A., P.G. and, V.G. were supported by Funda\c{c}\~{a}o para a Ci\^{e}ncia e a Tecnologia, with grants reference UIDB/00099/2020 and SFRH/BSAB/142940/2018.

\end{acknowledgements}

\bibliography{references}

\begin{thebibliography}{91}
\expandafter\ifx\csname natexlab\endcsname\relax\def\natexlab#1{#1}\fi

\bibitem[{{Abbott} {et~al.}(2016){Abbott}, {Abbott}, {Abbott}, {Abernathy},
  {Acernese}, {Ackley}, {Adams}, {Adams}, {Addesso}, {Adhikari}, \&
  et~al.}]{2016PhRvL.116f1102A}
{Abbott}, B.~P., {Abbott}, R., {Abbott}, T.~D., {et~al.} 2016, Physical Review
  Letters, 116, 061102

\bibitem[{Alexander(2005)}]{2005PhR...419...65A}
Alexander, T. 2005, Physics Reports, 419, 65

\bibitem[{{Alexander}(2017)}]{2017ARA&A..55...17A}
{Alexander}, T. 2017, \araa, 55, 17

\bibitem[{Ang{\'e}lil \& Saha(2014)}]{2014MNRAS.444.3780A}
Ang{\'e}lil, R. \& Saha, P. 2014, MNRAS, 444, 3780

\bibitem[{Ang{\'e}lil {et~al.}(2010)Ang{\'e}lil, Saha, \&
  Merritt}]{2010ApJ...720.1303A}
Ang{\'e}lil, R., Saha, P., \& Merritt, D. 2010, ApJ, 720, 1303

\bibitem[{{Bartko} {et~al.}(2009){Bartko}, {Martins}, {Fritz}, {Genzel},
  {Levin}, {Perets}, {Paumard}, {Nayakshin}, {Gerhard}, {Alexander},
  {Dodds-Eden}, {Eisenhauer}, {Gillessen}, {Mascetti}, {Ott}, {Perrin},
  {Pfuhl}, {Reid}, {Rouan}, {Sternberg}, \& {Trippe}}]{2009ApJ...697.1741B}
{Bartko}, H., {Martins}, F., {Fritz}, T.~K., {et~al.} 2009, \apj, 697, 1741

\bibitem[{{Baumgardt} {et~al.}(2018){Baumgardt}, {Amaro-Seoane}, \&
  {Sch{\"o}del}}]{2018A&A...609A..28B}
{Baumgardt}, H., {Amaro-Seoane}, P., \& {Sch{\"o}del}, R. 2018, \aap, 609, A28

\bibitem[{{Boehle} {et~al.}(2016){Boehle}, {Ghez}, {Sch{\"o}del}, {Meyer},
  {Yelda}, {Albers}, {Martinez}, {Becklin}, {Do}, {Lu}, {Matthews}, {Morris},
  {Sitarski}, \& {Witzel}}]{2016ApJ...830...17B}
{Boehle}, A., {Ghez}, A.~M., {Sch{\"o}del}, R., {et~al.} 2016, \apj, 830, 17

\bibitem[{Bonnet {et~al.}(2003)Bonnet, Str{\"o}bele, Biancat-Marchet, Brynnel,
  Conzelmann, Delabre, Donaldson, Farinato, Fedrigo, Hubin, Kasper, \&
  Kissler-Patig}]{2003SPIE.4839..329B}
Bonnet, H., Str{\"o}bele, S., Biancat-Marchet, F., {et~al.} 2003, Proc. SPIE
  Vol., 4839, 329

\bibitem[{{Cardoso} \& {Pani}(2019)}]{2019LRR....22....4C}
{Cardoso}, V. \& {Pani}, P. 2019, Living Reviews in Relativity, 22, 4

\bibitem[{{Chu} {et~al.}(2018){Chu}, {Do}, {Hees}, {Ghez}, {Naoz}, {Witzel},
  {Sakai}, {Chappell}, {Gautam}, {Lu}, \& {Matthews}}]{2018ApJ...854...12C}
{Chu}, D.~S., {Do}, T., {Hees}, A., {et~al.} 2018, \apj, 854, 12

\bibitem[{{Claesekens} \& {Hjort}(2008)}]{CH2008}
{Claesekens}, G. \& {Hjort}, N.~L. 2008, {Model Selection and Model Averaging}
  (Cambridge University Press)

\bibitem[{{Do} {et~al.}(2019){Do}, {Hees}, {Ghez}, {Martinez}, {Chu}, {Jia},
  {Sakai}, {Lu}, {Gautam}, {O'Neil}, {Becklin}, {Morris}, {Matthews},
  {Nishiyama}, {Campbell}, {Chappell}, {Chen}, {Ciurlo}, {Dehghanfar},
  {Gallego-Cano}, {Kerzendorf}, {Lyke}, {Naoz}, {Saida}, {Sch{\"o}del},
  {Takahashi}, {Takamori}, {Witzel}, \& {Wizinowich}}]{2019Sci...365..664D}
{Do}, T., {Hees}, A., {Ghez}, A., {et~al.} 2019, Science, 365, 664

\bibitem[{{Doeleman} {et~al.}(2008){Doeleman}, {Weintroub}, {Rogers},
  {Plambeck}, {Freund}, {Tilanus}, {Friberg}, {Ziurys}, {Moran}, {Corey},
  {Young}, {Smythe}, {Titus}, {Marrone}, {Cappallo}, {Bock}, {Bower},
  {Chamberlin}, {Davis}, {Krichbaum}, {Lamb}, {Maness}, {Niell}, {Roy},
  {Strittmatter}, {Werthimer}, {Whitney}, \& {Woody}}]{2008Natur.455...78D}
{Doeleman}, S.~S., {Weintroub}, J., {Rogers}, A.~E.~E., {et~al.} 2008, \nat,
  455, 78

\bibitem[{{Einstein}(1916)}]{1916AnP...354..769E}
{Einstein}, A. 1916, Annalen der Physik, 354, 769

\bibitem[{{Eisenhauer} {et~al.}(2003){Eisenhauer}, {Abuter}, {Bickert},
  {Biancat-Marchet}, {Bonnet}, {Brynnel}, {Conzelmann}, {Delabre}, {Donaldson},
  {Farinato}, {Fedrigo}, {Genzel}, {Hubin}, {Iserlohe}, {Kasper},
  {Kissler-Patig}, {Monnet}, {Roehrle}, {Schreiber}, {Stroebele}, {Tecza},
  {Thatte}, \& {Weisz}}]{2003SPIE.4841.1548E}
{Eisenhauer}, F., {Abuter}, R., {Bickert}, K., {et~al.} 2003, in \procspie,
  Vol. 4841, Instrument Design and Performance for Optical/Infrared
  Ground-based Telescopes, ed. M.~{Iye} \& A.~F.~M. {Moorwood}, 1548--1561

\bibitem[{Eisenhauer {et~al.}(2005)Eisenhauer, Genzel, Alexander, Abuter,
  Paumard, Ott, Gilbert, Gillessen, Horrobin, Trippe, Bonnet, Dumas, Hubin,
  Kaufer, Kissler-Patig, Monnet, Str{\"o}bele, Szeifert, Eckart, Sch{\"o}del,
  \& Zucker}]{2005ApJ...628..246E}
Eisenhauer, F., Genzel, R., Alexander, T., {et~al.} 2005, ApJ, 628, 246

\bibitem[{{Event Horizon Telescope Collaboration} {et~al.}(2019){Event Horizon
  Telescope Collaboration}, {Akiyama}, {Alberdi}, {Alef}, {Asada}, {Azulay},
  {Baczko}, {Ball}, {Balokovi{\'c}}, {Barrett}, {Bintley}, {Blackburn},
  {Boland}, {Bouman}, {Bower}, {Bremer}, {Brinkerink}, {Brissenden}, {Britzen},
  {Broderick}, {Broguiere}, {Bronzwaer}, {Byun}, {Carlstrom}, {Chael}, {Chan},
  {Chatterjee}, {Chatterjee}, {Chen}, {Chen}, {Cho}, {Christian}, {Conway},
  {Cordes}, {Crew}, {Cui}, {Davelaar}, {De Laurentis}, {Deane}, {Dempsey},
  {Desvignes}, {Dexter}, {Doeleman}, {Eatough}, {Falcke}, {Fish}, {Fomalont},
  {Fraga-Encinas}, {Freeman}, {Friberg}, {Fromm}, {G{\'o}mez}, {Galison},
  {Gammie}, {Garc{\'\i}a}, {Gentaz}, {Georgiev}, {Goddi}, {Gold}, {Gu},
  {Gurwell}, {Hada}, {Hecht}, {Hesper}, {Ho}, {Ho}, {Honma}, {Huang}, {Huang},
  {Hughes}, {Ikeda}, {Inoue}, {Issaoun}, {James}, {Jannuzi}, {Janssen},
  {Jeter}, {Jiang}, {Johnson}, {Jorstad}, {Jung}, {Karami}, {Karuppusamy},
  {Kawashima}, {Keating}, {Kettenis}, {Kim}, {Kim}, {Kim}, {Kino}, {Koay},
  {Koch}, {Koyama}, {Kramer}, {Kramer}, {Krichbaum}, {Kuo}, {Lauer}, {Lee},
  {Li}, {Li}, {Lindqvist}, {Liu}, {Liuzzo}, {Lo}, {Lobanov}, {Loinard},
  {Lonsdale}, {Lu}, {MacDonald}, {Mao}, {Markoff}, {Marrone}, {Marscher},
  {Mart{\'\i}-Vidal}, {Matsushita}, {Matthews}, {Medeiros}, {Menten}, {Mizuno},
  {Mizuno}, {Moran}, {Moriyama}, {Moscibrodzka}, {M{\"u}ller}, {Nagai},
  {Nagar}, {Nakamura}, {Narayan}, {Narayanan}, {Natarajan}, {Neri}, {Ni},
  {Noutsos}, {Okino}, {Olivares}, {Ortiz-Le{\'o}n}, {Oyama}, {{\"O}zel},
  {Palumbo}, {Patel}, {Pen}, {Pesce}, {Pi{\'e}tu}, {Plambeck}, {PopStefanija},
  {Porth}, {Prather}, {Preciado-L{\'o}pez}, {Psaltis}, {Pu}, {Ramakrishnan},
  {Rao}, {Rawlings}, {Raymond}, {Rezzolla}, {Ripperda}, {Roelofs}, {Rogers},
  {Ros}, {Rose}, {Roshanineshat}, {Rottmann}, {Roy}, {Ruszczyk}, {Ryan},
  {Rygl}, {S{\'a}nchez}, {S{\'a}nchez-Arguelles}, {Sasada}, {Savolainen},
  {Schloerb}, {Schuster}, {Shao}, {Shen}, {Small}, {Sohn}, {SooHoo}, {Tazaki},
  {Tiede}, {Tilanus}, {Titus}, {Toma}, {Torne}, {Trent}, {Trippe}, {Tsuda},
  {van Bemmel}, {van Langevelde}, {van Rossum}, {Wagner}, {Wardle},
  {Weintroub}, {Wex}, {Wharton}, {Wielgus}, {Wong}, {Wu}, {Young}, {Young},
  {Younsi}, {Yuan}, {Yuan}, {Zensus}, {Zhao}, {Zhao}, {Zhu}, {Algaba},
  {Allardi}, {Amestica}, {Anczarski}, {Bach}, {Baganoff}, {Beaudoin}, {Benson},
  {Berthold}, {Blanchard}, {Blundell}, {Bustamente}, {Cappallo},
  {Castillo-Dom{\'\i}nguez}, {Chang}, {Chang}, {Chang}, {Chen}, {Chilson},
  {Chuter}, {C{\'o}rdova Rosado}, {Coulson}, {Crawford}, {Crowley}, {David},
  {Derome}, {Dexter}, {Dornbusch}, {Dudevoir}, {Dzib}, {Eckart}, {Eckert},
  {Erickson}, {Everett}, {Faber}, {Farah}, {Fath}, {Folkers}, {Forbes},
  {Freund}, {G{\'o}mez-Ruiz}, {Gale}, {Gao}, {Geertsema}, {Graham}, {Greer},
  {Grosslein}, {Gueth}, {Haggard}, {Halverson}, {Han}, {Han}, {Hao},
  {Hasegawa}, {Henning}, {Hern{\'a}ndez-G{\'o}mez}, {Herrero-Illana},
  {Heyminck}, {Hirota}, {Hoge}, {Huang}, {Impellizzeri}, {Jiang}, {Kamble},
  {Keisler}, {Kimura}, {Kono}, {Kubo}, {Kuroda}, {Lacasse}, {Laing}, {Leitch},
  {Li}, {Lin}, {Liu}, {Liu}, {Lu}, {Marson}, {Martin-Cocher}, {Massingill},
  {Matulonis}, {McColl}, {McWhirter}, {Messias}, {Meyer-Zhao}, {Michalik},
  {Monta{\~n}a}, {Montgomerie}, {Mora-Klein}, {Muders}, {Nadolski}, {Navarro},
  {Neilsen}, {Nguyen}, {Nishioka}, {Norton}, {Nowak}, {Nystrom}, {Ogawa},
  {Oshiro}, {Oyama}, {Parsons}, {Paine}, {Pe{\~n}alver}, {Phillips}, {Poirier},
  {Pradel}, {Primiani}, {Raffin}, {Rahlin}, {Reiland}, {Risacher}, {Ruiz},
  {S{\'a}ez-Mada{\'\i}n}, {Sassella}, {Schellart}, {Shaw}, {Silva}, {Shiokawa},
  {Smith}, {Snow}, {Souccar}, {Sousa}, {Sridharan}, {Srinivasan}, {Stahm},
  {Stark}, {Story}, {Timmer}, {Vertatschitsch}, {Walther}, {Wei}, {Whitehorn},
  {Whitney}, {Woody}, {Wouterloot}, {Wright}, {Yamaguchi}, {Yu}, {Zeballos},
  {Zhang}, \& {Ziurys}}]{2019ApJ...875L...1E}
{Event Horizon Telescope Collaboration}, {Akiyama}, K., {Alberdi}, A., {et~al.}
  2019, \apjl, 875, L1

\bibitem[{{Fabian} {et~al.}(2000){Fabian}, {Iwasawa}, {Reynolds}, \&
  {Young}}]{2000PASP..112.1145F}
{Fabian}, A.~C., {Iwasawa}, K., {Reynolds}, C.~S., \& {Young}, A.~J. 2000,
  \pasp, 112, 1145

\bibitem[{{Falcke} \& {Markoff}(2013)}]{2013CQGra..30x4003F}
{Falcke}, H. \& {Markoff}, S.~B. 2013, Classical and Quantum Gravity, 30,
  244003

\bibitem[{{Falcke} {et~al.}(2000){Falcke}, {Melia}, \&
  {Agol}}]{2000ApJ...528L..13F}
{Falcke}, H., {Melia}, F., \& {Agol}, E. 2000, \apjl, 528, L13

\bibitem[{{Fritz} {et~al.}(2016){Fritz}, {Chatzopoulos}, {Gerhard},
  {Gillessen}, {Genzel}, {Pfuhl}, {Tacchella}, {Eisenhauer}, \&
  {Ott}}]{2016ApJ...821...44F}
{Fritz}, T.~K., {Chatzopoulos}, S., {Gerhard}, O., {et~al.} 2016, \apj, 821, 44

\bibitem[{{Gainutdinov}(2020)}]{2020arXiv200212598G}
{Gainutdinov}, R.~I. 2020, arXiv e-prints, arXiv:2002.12598

\bibitem[{Genzel {et~al.}(2010)Genzel, Eisenhauer, \&
  Gillessen}]{2010RvMP...82.3121G}
Genzel, R., Eisenhauer, F., \& Gillessen, S. 2010, Rev. Mod. Phys., 82, 3121

\bibitem[{Ghez {et~al.}(2003)Ghez, Duch{\^e}ne, Matthews, Hornstein, Tanner,
  Larkin, Morris, Becklin, Salim, Kremenek, Thompson, Soifer, Neugebauer, \&
  McLean}]{2003ApJ...586L.127G}
Ghez, A., Duch{\^e}ne, G., Matthews, K., {et~al.} 2003, ApJ Letters, 586, 127

\bibitem[{Ghez {et~al.}(2008)Ghez, Salim, Weinberg, Lu, Do, Dunn, Matthews,
  Morris, Yelda, Becklin, Kremenek, Milosavljevi{\'c}, \&
  Naiman}]{2008ApJ...689.1044G}
Ghez, A., Salim, S., Weinberg, N.~N., {et~al.} 2008, ApJ, 689, 1044

\bibitem[{{Gillessen} {et~al.}(2009{\natexlab{a}}){Gillessen}, {Eisenhauer},
  {Fritz}, {Bartko}, {Dodds-Eden}, {Pfuhl}, {Ott}, \&
  {Genzel}}]{2009ApJ...707L.114G}
{Gillessen}, S., {Eisenhauer}, F., {Fritz}, T.~K., {et~al.} 2009{\natexlab{a}},
  \apjl, 707, L114

\bibitem[{{Gillessen} {et~al.}(2009{\natexlab{b}}){Gillessen}, {Eisenhauer},
  {Trippe}, {Alexand er}, {Genzel}, {Martins}, \& {Ott}}]{2009ApJ...692.1075G}
{Gillessen}, S., {Eisenhauer}, F., {Trippe}, S., {et~al.} 2009{\natexlab{b}},
  \apj, 692, 1075

\bibitem[{{Gillessen} {et~al.}(2017){Gillessen}, {Plewa}, {Eisenhauer}, {Sari},
  {Waisberg}, {Habibi}, {Pfuhl}, {George}, {Dexter}, {von Fellenberg}, {Ott},
  \& {Genzel}}]{2017ApJ...837...30G}
{Gillessen}, S., {Plewa}, P.~M., {Eisenhauer}, F., {et~al.} 2017, \apj, 837, 30

\bibitem[{{Gondolo} \& {Silk}(1999)}]{1999PhRvL..83.1719G}
{Gondolo}, P. \& {Silk}, J. 1999, \prl, 83, 1719

\bibitem[{{Gravity Collaboration} {et~al.}(2017){Gravity Collaboration},
  {Abuter}, {Accardo}, {Amorim}, {Anugu}, {{\'A}vila}, {Azouaoui}, {Benisty},
  {Berger}, {Blind}, {Bonnet}, {Bourget}, {Brandner}, {Brast}, {Buron},
  {Burtscher}, {Cassaing}, {Chapron}, {Choquet}, {Cl{\'e}net}, {Collin},
  {Coud{\'e} Du Foresto}, {de Wit}, {de Zeeuw}, {Deen},
  {Delplancke-Str{\"o}bele}, {Dembet}, {Derie}, {Dexter}, {Duvert}, {Ebert},
  {Eckart}, {Eisenhauer}, {Esselborn}, {F{\'e}dou}, {Finger}, {Garcia}, {Garcia
  Dabo}, {Garcia Lopez}, {Gendron}, {Genzel}, {Gillessen}, {Gonte}, {Gordo},
  {Grould}, {Gr{\"o}zinger}, {Guieu}, {Haguenauer}, {Hans}, {Haubois}, {Haug},
  {Haussmann}, {Henning}, {Hippler}, {Horrobin}, {Huber}, {Hubert}, {Hubin},
  {Hummel}, {Jakob}, {Janssen}, {Jochum}, {Jocou}, {Kaufer}, {Kellner},
  {Kendrew}, {Kern}, {Kervella}, {Kiekebusch}, {Klein}, {Kok}, {Kolb}, {Kulas},
  {Lacour}, {Lapeyr{\`e}re}, {Lazareff}, {Le Bouquin}, {L{\`e}na}, {Lenzen},
  {L{\'e}v{\^e}que}, {Lippa}, {Magnard}, {Mehrgan}, {Mellein}, {M{\'e}rand},
  {Moreno-Ventas}, {Moulin}, {M{\"u}ller}, {M{\"u}ller}, {Neumann}, {Oberti},
  {Ott}, {Pallanca}, {Panduro}, {Pasquini}, {Paumard}, {Percheron}, {Perraut},
  {Perrin}, {Pfl{\"u}ger}, {Pfuhl}, {Phan Duc}, {Plewa}, {Popovic}, {Rabien},
  {Ram{\'{\i}}rez}, {Ramos}, {Rau}, {Riquelme}, {Rohloff}, {Rousset},
  {Sanchez-Bermudez}, {Scheithauer}, {Sch{\"o}ller}, {Schuhler}, {Spyromilio},
  {Straubmeier}, {Sturm}, {Suarez}, {Tristram}, {Ventura}, {Vincent},
  {Waisberg}, {Wank}, {Weber}, {Wieprecht}, {Wiest}, {Wiezorrek}, {Wittkowski},
  {Woillez}, {Wolff}, {Yazici}, {Ziegler}, \& {Zins}}]{2017A&A...602A..94G}
{Gravity Collaboration}, {Abuter}, R., {Accardo}, M., {et~al.} 2017, \aap, 602,
  A94

\bibitem[{{Gravity Collaboration} {et~al.}(2018{\natexlab{a}}){Gravity
  Collaboration}, {Abuter}, {Amorim}, {Anugu}, {Baub{\"o}ck}, {Benisty},
  {Berger}, {Blind}, {Bonnet}, {Brandner}, {Buron}, {Collin}, {Chapron},
  {Cl{\'e}net}, {Coud{\'e} Du Foresto}, {de Zeeuw}, {Deen},
  {Delplancke-Str{\"o}bele}, {Dembet}, {Dexter}, {Duvert}, {Eckart},
  {Eisenhauer}, {Finger}, {F{\"o}rster Schreiber}, {F{\'e}dou}, {Garcia},
  {Garcia Lopez}, {Gao}, {Gendron}, {Genzel}, {Gillessen}, {Gordo}, {Habibi},
  {Haubois}, {Haug}, {Hau{\ss}mann}, {Henning}, {Hippler}, {Horrobin},
  {Hubert}, {Hubin}, {Jimenez Rosales}, {Jochum}, {Jocou}, {Kaufer}, {Kellner},
  {Kendrew}, {Kervella}, {Kok}, {Kulas}, {Lacour}, {Lapeyr{\`e}re}, {Lazareff},
  {Le Bouquin}, {L{\'e}na}, {Lippa}, {Lenzen}, {M{\'e}rand}, {M{\"u}ler},
  {Neumann}, {Ott}, {Palanca}, {Paumard}, {Pasquini}, {Perraut}, {Perrin},
  {Pfuhl}, {Plewa}, {Rabien}, {Ram{\'{\i}}rez}, {Ramos}, {Rau},
  {Rodr{\'{\i}}guez-Coira}, {Rohloff}, {Rousset}, {Sanchez-Bermudez},
  {Scheithauer}, {Sch{\"o}ller}, {Schuler}, {Spyromilio}, {Straub},
  {Straubmeier}, {Sturm}, {Tacconi}, {Tristram}, {Vincent}, {von Fellenberg},
  {Wank}, {Waisberg}, {Widmann}, {Wieprecht}, {Wiest}, {Wiezorrek}, {Woillez},
  {Yazici}, {Ziegler}, \& {Zins}}]{2018A&A...615L..15G}
{Gravity Collaboration}, {Abuter}, R., {Amorim}, A., {et~al.}
  2018{\natexlab{a}}, \aap, 615, L15

\bibitem[{{Gravity Collaboration} {et~al.}(2019){Gravity Collaboration},
  {Abuter}, {Amorim}, {Baub{\"o}ck}, {Berger}, {Bonnet}, {Brand ner},
  {Cl{\'e}net}, {Coud{\'e} Du Foresto}, {de Zeeuw}, {Dexter}, {Duvert},
  {Eckart}, {Eisenhauer}, {F{\"o}rster Schreiber}, {Garcia}, {Gao}, {Gendron},
  {Genzel}, {Gerhard}, {Gillessen}, {Habibi}, {Haubois}, {Henning}, {Hippler},
  {Horrobin}, {Jim{\'e}nez-Rosales}, {Jocou}, {Kervella}, {Lacour},
  {Lapeyr{\`e}re}, {Le Bouquin}, {L{\'e}na}, {Ott}, {Paumard}, {Perraut},
  {Perrin}, {Pfuhl}, {Rabien}, {Rodriguez Coira}, {Rousset}, {Scheithauer},
  {Sternberg}, {Straub}, {Straubmeier}, {Sturm}, {Tacconi}, {Vincent}, {von
  Fellenberg}, {Waisberg}, {Widmann}, {Wieprecht}, {Wiezorrek}, {Woillez}, \&
  {Yazici}}]{2019A&A...625L..10G}
{Gravity Collaboration}, {Abuter}, R., {Amorim}, A., {et~al.} 2019, \aap, 625,
  L10

\bibitem[{{Gravity Collaboration} {et~al.}(2018{\natexlab{b}}){Gravity
  Collaboration}, {Abuter}, {Amorim}, {Baub{\"o}ck}, {Berger}, {Bonnet},
  {Brandner}, {Cl{\'e}net}, {Coud{\'e} Du Foresto}, {de Zeeuw}, {Deen},
  {Dexter}, {Duvert}, {Eckart}, {Eisenhauer}, {F{\"o}rster Schreiber},
  {Garcia}, {Gao}, {Gendron}, {Genzel}, {Gillessen}, {Guajardo}, {Habibi},
  {Haubois}, {Henning}, {Hippler}, {Horrobin}, {Huber}, {Jim{\'e}nez-Rosales},
  {Jocou}, {Kervella}, {Lacour}, {Lapeyr{\`e}re}, {Lazareff}, {Le Bouquin},
  {L{\'e}na}, {Lippa}, {Ott}, {Panduro}, {Paumard}, {Perraut}, {Perrin},
  {Pfuhl}, {Plewa}, {Rabien}, {Rodr{\'{\i}}guez-Coira}, {Rousset}, {Sternberg},
  {Straub}, {Straubmeier}, {Sturm}, {Tacconi}, {Vincent}, {von Fellenberg},
  {Waisberg}, {Widmann}, {Wieprecht}, {Wiezorrek}, {Woillez}, \&
  {Yazici}}]{2018A&A...618L..10G}
{Gravity Collaboration}, {Abuter}, R., {Amorim}, A., {et~al.}
  2018{\natexlab{b}}, \aap, 618, L10

\bibitem[{{Grould} {et~al.}(2017){Grould}, {Vincent}, {Paumard}, \&
  {Perrin}}]{2017A&A...608A..60G}
{Grould}, M., {Vincent}, F.~H., {Paumard}, T., \& {Perrin}, G. 2017, \aap, 608,
  A60

\bibitem[{{Gualandris} {et~al.}(2010){Gualandris}, {Gillessen}, \&
  {Merritt}}]{2010MNRAS.409.1146G}
{Gualandris}, A., {Gillessen}, S., \& {Merritt}, D. 2010, \mnras, 409, 1146

\bibitem[{Gualandris \& Merritt(2009)}]{2009ApJ...705..361G}
Gualandris, A. \& Merritt, D. 2009, ApJ, 705, 361

\bibitem[{{Habibi} {et~al.}(2017){Habibi}, {Gillessen}, {Martins},
  {Eisenhauer}, {Plewa}, {Pfuhl}, {George}, {Dexter}, {Waisberg}, {Ott}, {von
  Fellenberg}, {Baub{\"o}ck}, {Jimenez-Rosales}, \&
  {Genzel}}]{2017ApJ...847..120H}
{Habibi}, M., {Gillessen}, S., {Martins}, F., {et~al.} 2017, \apj, 847, 120

\bibitem[{{Hansen} \& {Milosavljevi{\'c}}(2003)}]{2003ApJ...593L..77H}
{Hansen}, B. M.~S. \& {Milosavljevi{\'c}}, M. 2003, \apjl, 593, L77

\bibitem[{{Hees} {et~al.}(2017){Hees}, {Do}, {Ghez}, {Martinez}, {Naoz},
  {Becklin}, {Boehle}, {Chappell}, {Chu}, {Dehghanfar}, {Kosmo}, {Lu},
  {Matthews}, {Morris}, {Sakai}, {Sch{\"o}del}, \&
  {Witzel}}]{2017PhRvL.118u1101H}
{Hees}, A., {Do}, T., {Ghez}, A.~M., {et~al.} 2017, Physical Review Letters,
  118, 211101

\bibitem[{{Hofmann} {et~al.}(1993){Hofmann}, {Eckart}, {Genzel}, \&
  {Drapatz}}]{1993Ap&SS.205....1H}
{Hofmann}, R., {Eckart}, A., {Genzel}, R., \& {Drapatz}, S. 1993, \apss, 205, 1

\bibitem[{{Issaoun} {et~al.}(2019){Issaoun}, {Johnson}, {Blackburn},
  {Brinkerink}, {Mo{\'s}cibrodzka}, {Chael}, {Goddi}, {Mart{\'\i}-Vidal},
  {Wagner}, {Doeleman}, {Falcke}, {Krichbaum}, {Akiyama}, {Bach}, {Bouman},
  {Bower}, {Broderick}, {Cho}, {Crew}, {Dexter}, {Fish}, {Gold}, {G{\'o}mez},
  {Hada}, {Hern{\'a}ndez-G{\'o}mez}, {Jan{\ss}en}, {Kino}, {Kramer}, {Loinard},
  {Lu}, {Markoff}, {Marrone}, {Matthews}, {Moran}, {M{\"u}ller}, {Roelofs},
  {Ros}, {Rottmann}, {Sanchez}, {Tilanus}, {de Vicente}, {Wielgus}, {Zensus},
  \& {Zhao}}]{2019ApJ...871...30I}
{Issaoun}, S., {Johnson}, M.~D., {Blackburn}, L., {et~al.} 2019, \apj, 871, 30

\bibitem[{{Johnson} {et~al.}(2017){Johnson}, {Bouman}, {Blackburn}, {Chael},
  {Rosen}, {Shiokawa}, {Roelofs}, {Akiyama}, {Fish}, \&
  {Doeleman}}]{2017ApJ...850..172J}
{Johnson}, M.~D., {Bouman}, K.~L., {Blackburn}, L., {et~al.} 2017, \apj, 850,
  172

\bibitem[{{Kerr}(1965)}]{1965qssg.conf...99K}
{Kerr}, R.~P. 1965, in Quasi-Stellar Sources and Gravitational Collapse, ed.
  I.~{Robinson}, A.~{Schild}, \& E.~L. {Schucking}, 99

\bibitem[{{Kormendy} \& {Ho}(2013)}]{2013ARA&A..51..511K}
{Kormendy}, J. \& {Ho}, L.~C. 2013, \araa, 51, 511

\bibitem[{{Kramer}(2016)}]{2016IJMPD..2530029K}
{Kramer}, M. 2016, International Journal of Modern Physics D, 25, 1630029

\bibitem[{{Kramer} {et~al.}(2006){Kramer}, {Stairs}, {Manchester},
  {McLaughlin}, {Lyne}, {Ferdman}, {Burgay}, {Lorimer}, {Possenti}, {D'Amico},
  {Sarkissian}, {Hobbs}, {Reynolds}, {Freire}, \&
  {Camilo}}]{2006Sci...314...97K}
{Kramer}, M., {Stairs}, I.~H., {Manchester}, R.~N., {et~al.} 2006, Science,
  314, 97

\bibitem[{{Lacroix}(2018)}]{2018A&A...619A..46L}
{Lacroix}, T. 2018, \aap, 619, A46

\bibitem[{Lenzen {et~al.}(1998)Lenzen, Hofmann, Bizenberger, \&
  Tusche}]{1998SPIE.3354..606L}
Lenzen, R., Hofmann, R., Bizenberger, P., \& Tusche, A. 1998, Proc. SPIE Vol.,
  3354, 606

\bibitem[{{Levenberg}(1944)}]{1944L}
{Levenberg}, K. 1944, Quart. Appl. Math, 2, 164

\bibitem[{{Lynden-Bell} \& {Rees}(1971)}]{1971MNRAS.152..461L}
{Lynden-Bell}, D. \& {Rees}, M.~J. 1971, \mnras, 152, 461

\bibitem[{{Marquardt}(1963)}]{1963M}
{Marquardt}, D.~W. 1963, J. Soc. Indust. Appl. Math,, 11, 431

\bibitem[{Martins {et~al.}(2008)Martins, Gillessen, Eisenhauer, Genzel, Ott, \&
  Trippe}]{2008ApJ...672L.119M}
Martins, F., Gillessen, S., Eisenhauer, F., {et~al.} 2008, ApJ Letters, 672,
  119

\bibitem[{{McConnell} \& {Ma}(2013)}]{2013ApJ...764..184M}
{McConnell}, N.~J. \& {Ma}, C.-P. 2013, \apj, 764, 184

\bibitem[{{Merritt} {et~al.}(2010){Merritt}, {Alexander}, {Mikkola}, \&
  {Will}}]{2010PhRvD..81f2002M}
{Merritt}, D., {Alexander}, T., {Mikkola}, S., \& {Will}, C.~M. 2010, \prd, 81,
  062002

\bibitem[{{Merritt} {et~al.}(2009){Merritt}, {Gualandris}, \&
  {Mikkola}}]{2009ApJ...693L..35M}
{Merritt}, D., {Gualandris}, A., \& {Mikkola}, S. 2009, \apjl, 693, L35

\bibitem[{Meyer {et~al.}(2012)Meyer, Ghez, Sch{\"o}del, Yelda, Boehle, Lu, Do,
  Morris, Becklin, \& Matthews}]{2012Sci...338...84M}
Meyer, L., Ghez, A., Sch{\"o}del, R., {et~al.} 2012, Science, 338, 84

\bibitem[{{Misner} {et~al.}(1973){Misner}, {Thorne}, \&
  {Wheeler}}]{1973grav.book.....M}
{Misner}, C.~W., {Thorne}, K.~S., \& {Wheeler}, J.~A. 1973, {Gravitation} (W.H.
  Freeman, Princeton Univers. Press)

\bibitem[{{Moran} {et~al.}(1999){Moran}, {Greenhill}, \&
  {Herrnstein}}]{1999JApA...20..165M}
{Moran}, J.~M., {Greenhill}, L.~J., \& {Herrnstein}, J.~R. 1999, Journal of
  Astrophysics and Astronomy, 20, 165

\bibitem[{{Morris} {et~al.}(2012){Morris}, {Meyer}, \&
  {Ghez}}]{2012RAA....12..995M}
{Morris}, M.~R., {Meyer}, L., \& {Ghez}, A.~M. 2012, Research in Astronomy and
  Astrophysics, 12, 995

\bibitem[{{Mouawad} {et~al.}(2005){Mouawad}, {Eckart}, {Pfalzner},
  {Sch{\"o}del}, {Moultaka}, \& {Spurzem}}]{2005AN....326...83M}
{Mouawad}, N., {Eckart}, A., {Pfalzner}, S., {et~al.} 2005, Astronomische
  Nachrichten, 326, 83

\bibitem[{{Naoz} {et~al.}(2020){Naoz}, {Will}, {Ramirez-Ruiz}, {Hees}, {Ghez},
  \& {Do}}]{2020ApJ...888L...8N}
{Naoz}, S., {Will}, C.~M., {Ramirez-Ruiz}, E., {et~al.} 2020, \apjl, 888, L8

\bibitem[{{Navarro} {et~al.}(1996){Navarro}, {Frenk}, \&
  {White}}]{1996ApJ...462..563N}
{Navarro}, J.~F., {Frenk}, C.~S., \& {White}, S. D.~M. 1996, \apj, 462, 563

\bibitem[{{Parsa} {et~al.}(2017){Parsa}, {Eckart}, {Shahzamanian}, {Karas},
  {Zaja{\v c}ek}, {Zensus}, \& {Straubmeier}}]{2017ApJ...845...22P}
{Parsa}, M., {Eckart}, A., {Shahzamanian}, B., {et~al.} 2017, \apj, 845, 22

\bibitem[{{Paumard} {et~al.}(2006){Paumard}, {Genzel}, {Martins}, {Nayakshin},
  {Beloborodov}, {Levin}, {Trippe}, {Eisenhauer}, {Ott}, {Gillessen}, {Abuter},
  {Cuadra}, {Alexander}, \& {Sternberg}}]{2006ApJ...643.1011P}
{Paumard}, T., {Genzel}, R., {Martins}, F., {et~al.} 2006, \apj, 643, 1011

\bibitem[{{Perrin} \& {Woillez}(2019)}]{2019A&A...625A..48P}
{Perrin}, G. \& {Woillez}, J. 2019, \aap, 625, A48

\bibitem[{Plewa {et~al.}(2015)Plewa, Gillessen, Eisenhauer, Ott, Pfuhl, George,
  Dexter, Habibi, Genzel, Reid, \& Menten}]{2015MNRAS.453.3234P}
Plewa, P.~M., Gillessen, S., Eisenhauer, F., {et~al.} 2015, MNRAS, 453, 3234

\bibitem[{{Plewa} \& {Sari}(2018)}]{2018MNRAS.476.4372P}
{Plewa}, P.~M. \& {Sari}, R. 2018, \mnras, 476, 4372

\bibitem[{{Plummer}(1911)}]{1911MNRAS..71..460P}
{Plummer}, H.~C. 1911, \mnras, 71, 460

\bibitem[{{Psaltis}(2004)}]{2004AIPC..714...29P}
{Psaltis}, D. 2004, in American Institute of Physics Conference Series, Vol.
  714, X-ray Timing 2003: Rossi and Beyond, ed. P.~{Kaaret}, F.~K. {Lamb}, \&
  J.~H. {Swank}, 29--35

\bibitem[{{Psaltis} {et~al.}(2016){Psaltis}, {Wex}, \&
  {Kramer}}]{2016ApJ...818..121P}
{Psaltis}, D., {Wex}, N., \& {Kramer}, M. 2016, \apj, 818, 121

\bibitem[{Reid \& Brunthaler(2004)}]{2004ApJ...616..872R}
Reid, M.~J. \& Brunthaler, A. 2004, ApJ, 616, 872

\bibitem[{{Reid} \& {Brunthaler}(2020)}]{2020arXiv200104386R}
{Reid}, M.~J. \& {Brunthaler}, A. 2020, arXiv e-prints, arXiv:2001.04386

\bibitem[{Reid {et~al.}(2009)Reid, Menten, Zheng, Brunthaler, \&
  Xu}]{Reid:2009p2548}
Reid, M.~J., Menten, K.~M., Zheng, X.~W., Brunthaler, A., \& Xu, Y. 2009, ApJ,
  705, 1548

\bibitem[{Rousset {et~al.}(1998)Rousset, Lacombe, Puget, Hubin, Gendron, Conan,
  Kern, Madec, Rabaud, Mouillet, Lagrange, \& Rigaut}]{1998SPIE.3353..508R}
Rousset, G., Lacombe, F., Puget, P., {et~al.} 1998, Proc. SPIE Vol., 3353, 508

\bibitem[{{Rubilar} \& {Eckart}(2001)}]{2001A&A...374...95R}
{Rubilar}, G.~F. \& {Eckart}, A. 2001, \aap, 374, 95

\bibitem[{{Sakai} {et~al.}(2019){Sakai}, {Lu}, {Ghez}, {Jia}, {Do}, {Witzel},
  {Gautam}, {Hees}, {Becklin}, {Matthews}, \& {Hosek}}]{2019ApJ...873...65S}
{Sakai}, S., {Lu}, J.~R., {Ghez}, A., {et~al.} 2019, \apj, 873, 65

\bibitem[{{Schmidt}(1963)}]{1963Natur.197.1040S}
{Schmidt}, M. 1963, \nat, 197, 1040

\bibitem[{{Sch{\"o}del} {et~al.}(2018){Sch{\"o}del}, {Gallego-Cano}, {Dong},
  {Nogueras-Lara}, {Gallego-Calvente}, {Amaro-Seoane}, \&
  {Baumgardt}}]{2018A&A...609A..27S}
{Sch{\"o}del}, R., {Gallego-Cano}, E., {Dong}, H., {et~al.} 2018, \aap, 609,
  A27

\bibitem[{Sch{\"o}del {et~al.}(2009)Sch{\"o}del, Merritt, \&
  Eckart}]{2009A&A...502...91S}
Sch{\"o}del, R., Merritt, D., \& Eckart, A. 2009, A{\&}A, 502, 91

\bibitem[{Sch{\"o}del {et~al.}(2002)Sch{\"o}del, Ott, Genzel, Hofmann, Lehnert,
  Eckart, Mouawad, Alexander, Reid, Lenzen, Hartung, Lacombe, Rouan, Gendron,
  Rousset, Lagrange, Brandner, Ageorges, Lidman, Moorwood, Spyromilio, Hubin,
  \& Menten}]{2002Natur.419..694S}
Sch{\"o}del, R., Ott, T., Genzel, R., {et~al.} 2002, Nature, 419, 694

\bibitem[{{Schwarzschild}(1916)}]{1916SPAW.......189S}
{Schwarzschild}, K. 1916, Sitzungsberichte der K{\"o}niglich Preu{\ss}ischen
  Akademie der Wissenschaften (Berlin), 189

\bibitem[{{Tanaka} {et~al.}(1995){Tanaka}, {Nandra}, {Fabian}, {Inoue},
  {Otani}, {Dotani}, {Hayashida}, {Iwasawa}, {Kii}, {Kunieda}, {Makino}, \&
  {Matsuoka}}]{1995Natur.375..659T}
{Tanaka}, Y., {Nandra}, K., {Fabian}, A.~C., {et~al.} 1995, \nat, 375, 659

\bibitem[{{Vincent} {et~al.}(2011){Vincent}, {Paumard}, {Gourgoulhon}, \&
  {Perrin}}]{2011CQGra..28v5011V}
{Vincent}, F.~H., {Paumard}, T., {Gourgoulhon}, E., \& {Perrin}, G. 2011,
  Classical and Quantum Gravity, 28, 225011

\bibitem[{{Will}(2008)}]{2008ApJ...674L..25W}
{Will}, C.~M. 2008, \apjl, 674, L25

\bibitem[{{Will}(2014)}]{2014LRR....17....4W}
{Will}, C.~M. 2014, Living Reviews in Relativity, 17, 4

\bibitem[{{Will} \& {Nordtvedt}(1972)}]{1972ApJ...177..757W}
{Will}, C.~M. \& {Nordtvedt}, Kenneth, J. 1972, \apj, 177, 757

\bibitem[{{Yelda} {et~al.}(2014){Yelda}, {Ghez}, {Lu}, {Do}, {Meyer}, {Morris},
  \& {Matthews}}]{2014ApJ...783..131Y}
{Yelda}, S., {Ghez}, A.~M., {Lu}, J.~R., {et~al.} 2014, \apj, 783, 131

\bibitem[{{Yu} \& {Tremaine}(2003)}]{2003ApJ...599.1129Y}
{Yu}, Q. \& {Tremaine}, S. 2003, \apj, 599, 1129

\bibitem[{{Zhang} \& {Iorio}(2017)}]{2017ApJ...834..198Z}
{Zhang}, F. \& {Iorio}, L. 2017, \apj, 834, 198

\bibitem[{Zucker {et~al.}(2006)Zucker, Alexander, Gillessen, Eisenhauer, \&
  Genzel}]{2006ApJ...639L..21Z}
Zucker, S., Alexander, T., Gillessen, S., Eisenhauer, F., \& Genzel, R. 2006,
  ApJ Letters, 639, 21

\end{thebibliography}

\begin{appendix}

\section{Experimental techniques }
  \label{sec:A}

\subsection{GRAVITY data analysis}
\label{sec:A1}

Our result crucially depends on the use of GRAVITY, the VLTI beam combiner, which as a result of its extremely high angular resolution of $\approx 3\,$mas yields very accurate astrometry with errors well below $100\,\mu$as (\citealt{2017A&A...602A..94G} and Figure~\ref{fig:figA3}). Depending on the separation between S2 and Sgr~A* there are two fundamentally different ways to retrieve the separation vector between S2 and Sgr~A*.

 \begin{figure}[h]
\centering
\includegraphics[width=7cm]{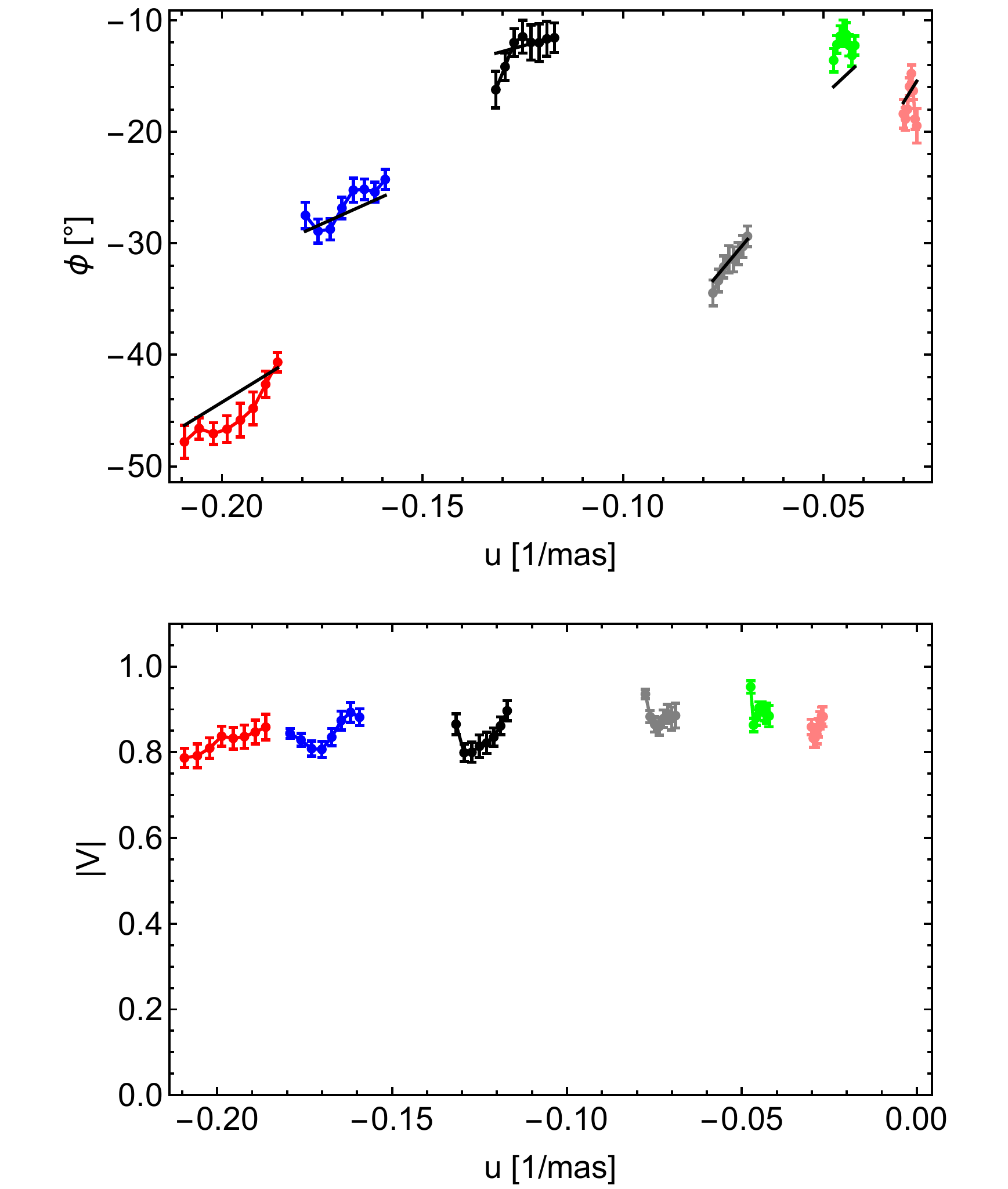}
\caption{Example of a unary fit for a five-minute exposure on Sgr~A* from April 22, 2019, 06:39:55. Top: Phase $\Phi$ as a function of projected baseline vector $\vec{u}$ for the s-polarisation channel. Per baseline, eight spectral channels were included here. The black line is a unary model, which yields the offset from the interferometer pointing position to $\Delta \mathrm{RA} = 322 \pm 9\,\mu$as and $\Delta \mathrm{Dec} = 301 \pm 8\,\mu$as (formal fit errors). Bottom: The visibility modulus for the same data is constant and close to unity, consistent with the choice of fitting a single point source.}
\label{fig:figA1}
\end{figure}

{\bf Dual-beam method.} For separations larger than the single-telescope beam size (FWHM$\,\approx  60\,$mas), the GRAVITY science channel fibre needs to be pointed once to Sgr~A* and once to S2, such that the respective target is the dominant source in the field. The phases of the complex visibilities of each pointing then yield an accurate distance to the fringe-tracking star, IRS16C in our case. By interferometrically calibrating the Sgr~A* data with S2, the position of IRS16C drops out, and we obtain a data set in which the six phases $\Phi_i$ directly measure the desired separation vector $\vec{s} = (\Delta\mathrm{RA}, \Delta\mathrm{Dec})$ between S2 and Sgr~A* through the basic interferometer formula for a unary model,
\begin{equation}
\Phi_{i,j}= 2 \pi \, \vec{s} \cdot \vec{B}_i /\lambda_j\,\,,
\end{equation}
where $\vec{B}_i$ denotes the $i$-th of the six baselines. Because our data are spectrally resolved into 14 channels $\lambda_j$ across the K band ($2.0\,\mu$m to $2.4\,\mu$m), the unknown \vec{s} with two parameters is well constrained by a fit to the phases.
This method applies mostly to the 2019 data, and partly to the 2017 data. In Figure~\ref{fig:figA1} we show an example for such a unary fit for one Sgr~A* exposure.

 \begin{figure}[h]
\centering
\includegraphics[width=6cm]{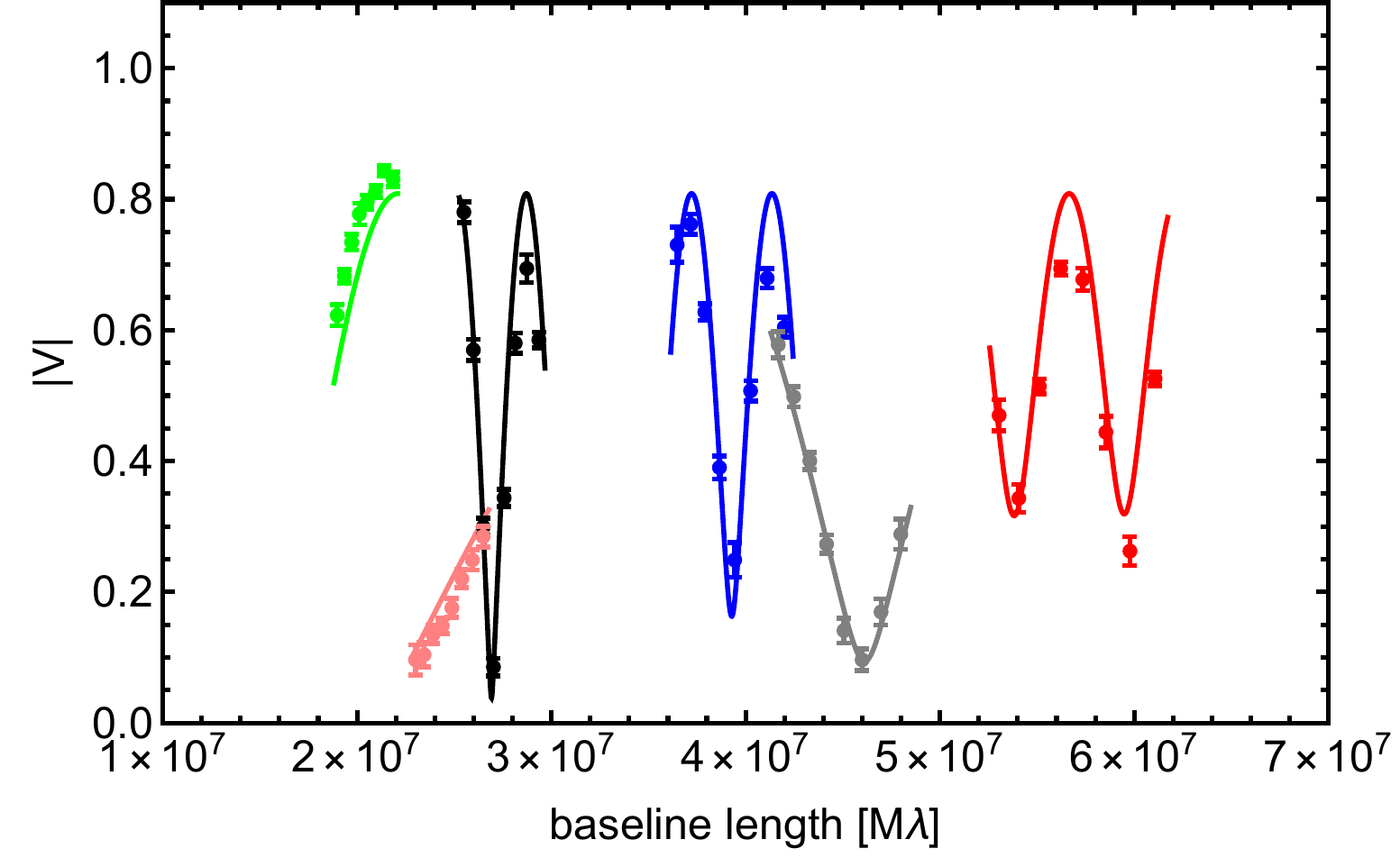}
\caption{Example of a binary fit from July 8, 2017, 03:27:51. The observed visibility modulus shows strong modulation, the signature of an interferometrically resolved binary. The lines show the model for the six baselines, which includes the separation vector. The formal uncertainty in this example fit is $8\,\mu$as per coordinate.}
\label{fig:figA2}
\end{figure}

\begin{figure*}[t!]
\centering
\begin{minipage}{.72\textwidth}
\includegraphics[width=\linewidth]{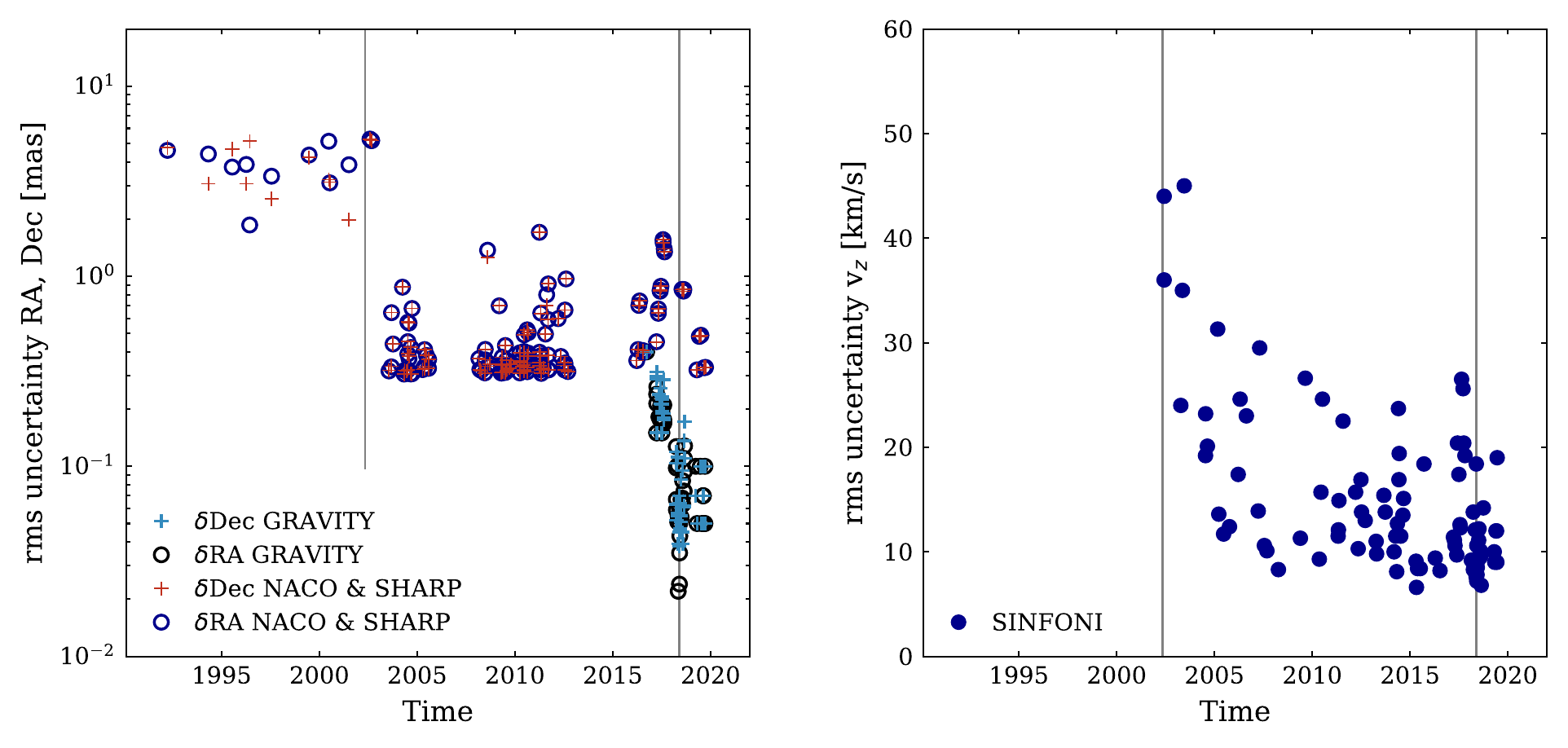}
\end{minipage} \quad
\begin{minipage}{.25\textwidth}
\caption{Astrometric (left) and spectroscopic (right) $1\sigma$ statistical measurement uncertainties of S2 over time. The left panel shows the almost 100-fold improvement in astrometric precision in RA and Dec, from the early period of speckle imagery with SHARP on the $3.5\,$m NTT (until 2001), then the AO imagery with NACO on the $8\,m$ VLT (>2002), and then, since 2016.7, the interferometric astrometry of GRAVITY combining all four $8\,$m telescopes of the VLT. The grey vertical lines are the two pericentre passages (2002.33 and 2018.38) covered by our data set.}
\label{fig:figA3}
\end{minipage} 
\end{figure*}

{\bf Single-beam method.} For separations below the single-telescope beam size, both sources are observed simultaneously and appear as an interferometric binary. In this case, the amplitudes of the complex visibilities as well as the closure phases carry the signature, which is a beating pattern in each baseline along the spectral axis. We fitted a binary model to these data, for which the complex visibilities are:
\begin{equation}
\mathbb{C}_{k,l} = \frac{I_E + \sqrt{f_k f_l} I_C}{\sqrt{I_A +f_k I_B+f_\mathrm{BG} I_D}\,\,\sqrt{I_A +f_l I_B+f_\mathrm{BG} I_D}}\,\,.
\label{eqA1}
\end{equation}
In this expression we use the abbreviations
\begin{eqnarray}
I_A &=& I(\alpha_\mathrm{Sgr},0) \,\,,\nonumber \\
I_B &=& I(\alpha_\mathrm{S2},0) \,\,,\nonumber \\
I_C &=& I(\alpha_\mathrm{S2},\mathrm{OPD}_\mathrm{S2}) \,\,,\nonumber \\
I_D &=& I(\alpha_\mathrm{BG},0)\,\,, \nonumber \\
I_E &=& I(\alpha_\mathrm{Sgr},\mathrm{OPD}_\mathrm{Sgr}) \,\,,\nonumber 
\end{eqnarray}
where
\begin{equation}
I(\alpha,\mathrm{OPD}) = \int_{\Delta \lambda}  P(\lambda) \,\lambda^{-1-\alpha}_{2.2}\, e^{-2\pi i \, \mathrm{OPD}/\lambda} d\lambda\,\,.
\label{eqA2}
\end{equation}
The $\alpha$ are the spectral slopes of Sgr~A*, S2, and background, and $\lambda_{2.2}$ is the wavelength $\lambda$ divided by the reference wavelength $\lambda_0=2.2\,\mu$m. The optical path differences for X$\,=\,$S2 and X$\,=\,$Sgr~A* are
\begin{equation}
\mathrm{OPD_X} = \vec{s_\mathrm{X}}\cdot \vec{B}_{k,l}\,\,.
\end{equation}
The function $P(\lambda)$ is the spectral bandpass, for which we used a top-hat function with a width corresponding to the measured spectral resolution. The $f_k$ and $f_l$ are the flux ratios of S2 to Sgr~A* for telescope $k$ and $l$; $f_\mathrm{BG}$ is the flux ratio of unresolved background to the Sgr~A* flux. The model yields  a complex visibility for all baselines and spectral channels, of which we fit the amplitudes and closure phases to the data.
We also used this analysis in our previous work \citep{2018A&A...615L..15G, 2018A&A...618L..10G} and here for the 2018 and  2017 data. In Figure~\ref{fig:figA2} we show an example of how the binary model describes the visibility amplitudes for one exposure.

\subsection{Details of the unary model fits}
\label{sec:A2} 
The aim is to measure the separation vector between S2 and Sgr~A*. GRAVITY measures the separation between science object and fringe-tracking star (IRS16C in our case). The desired separation is obtained by measuring both S2 and Sgr~A* with respect to IRS16C, and subtracting the two measurements. This corresponds to interferometrically calibrating the phases of Sgr~A* with those of S2. 

By construction, the phases of the calibrator S2 frame are identical to 0, and ideally, the phases for all other S2 frames are 0 as well. In reality, this is not the case. At the time of observing, the separation vector $\vec{r}$ between the fringe-tracking star IRS16C and S2 needs to be provided to GRAVITY for tracking the fringes with the differential delay lines. At this point, $\vec{r}$ is not known to the interferometric precision, but only from the AO-data based orbital motion of IRS16C \citep{2017ApJ...837...30G}. For a subsequent S2 file, when the projected baselines have changed by some value $\Delta\vec{B}$ due to Earth rotation, the error in pointing $\Delta \vec{r}$ therefore leads to an additional phase $\Delta\Phi = \Delta\vec{B} \cdot  \Delta \vec{r}$. By observing S2 a few times per night, we obtain a set of constraints for  $\Delta \vec{r}$, which allows fitting for  $\Delta \vec{r}$ over the course of the night. 

Therefore we can correct our data post-facto for this offset $\Delta \vec{r}$, and obtain phases for Sgr~A* that directly relate to S2. Because S2 is several interferometric beams away from Sgr~A*, the phases are still wrapped, which is inconvenient for fitting. The solution is to subtract the separation vector $\vec{r}$ as provided at the time of observing, and only fit (using the  $\Delta \vec{r}$-corrected phases) the difference to that separation.

The choice of which of the S2 frames we use as calibrator depends on the night and on the data quality of the individual files. Ideally, we seek S2 frames of good quality close in time to the Sgr~A* frames, in which Sgr~A* was bright. Typically, the Sgr~A* frames during which the source  is clearly detectable (flares of at least moderate brightness with $m_K < 16$) lead to a well-determined and stable S2-Sgr~A* vector.

\subsection{Details of the binary model fits}
\label{sec:A3}
We used the binary fitting method in our previous publications \citep{2018A&A...615L..15G, 2018A&A...618L..10G, 2019A&A...625L..10G}. The quantities used are the visibility amplitudes and the closure phases, both of which measure the internal source structure. We omit the visibility phases here, because they mostly contain information about the location of the phase centre and only to a lesser degree about the source internal structure. One of the parameters describing the source structure is the desired binary separation.

Here, we also correct for static aberrations during the binary fitting, refining our earlier procedure as a result of an improved understanding of the instrumental systematics. The aberrations are induced in the fibre coupling unit of GRAVITY and distort the incoming wavefronts depending on the source's position in the field of view (FOV). The effect is zero at the centre of the FOV but increases with off-axis distance and thus is of particular importance for the 2017 data where S2 and Sgr~A* are detected simultaneously in a single fibre-beam positioning at a separation comparable to the fibre FOV.

We parametrise the effect of a static aberration with an amplitude $A^\mathrm{off}$ and a phase $\Phi^\mathrm{off}$ on a plain wavefront in complex notation as
\begin{equation}
\Psi_k = E_0\, A_k^\mathrm{off} \exp(i\omega t + i \,\vec{s} \cdot \vec{x_k} + i \,\Phi_k^\mathrm{off}) \,\,,
\end{equation}
where $k$ labels the telescope and $\vec{x_k}$ denotes its position, $E_0$ is the amplitude of the unperturbed electric field, and $\vec{s}$ is the source position on the sky. The scaling in amplitude $A_k^\mathrm{off}$ and the phase shift $\Phi_k^\mathrm{off}$ are functions of the source position with respect to the field centre and differ for each telescope. 

The GRAVITY pipeline determines the normalised interferometric visibility from the correlated flux of two telescopes divided by their respective individual fluxes. The field-dependent aberrations enter the Van Cittert-Zernike theorem as
 \begin{equation}
V_{kl}=\frac{\int I(\vec{\sigma})A_k^\mathrm{off}(\vec{\sigma}) A_l^\mathrm{off}(\vec{\sigma}) 
e^{\frac{2\pi i}{\lambda}\vec{\sigma} \cdot \vec{b}_{kl}+ i \Phi_k^\mathrm{off}(\vec{\sigma})-i \Phi_l^\mathrm{off}(\vec{\sigma}) } d\vec{\sigma}}
{\sqrt{\int I(\vec{\sigma}) \left( A_k^\mathrm{off}(\vec{\sigma}) \right)^2 d\vec{\sigma} \times \int I(\vec{\sigma}) \left( A_l^\mathrm{off}(\vec{\sigma}) \right)^2 d\vec{\sigma} }} \,\,,
\end{equation}
where $I(\vec{\sigma})$ is the source intensity distribution and $\vec{b}_{kl}$ is the projection of the baseline vector onto the plane perpendicular to the line of sight. The expression for a binary system follows from this equation and generalises Eq.~\ref{eqA1}. The integrals in Eq.~\ref{eqA2}  then read as 
\begin{equation}
I(\alpha,\vec{s}) = \int_{\Delta \lambda}  P(\lambda) \,\lambda^{-1-\alpha}_{2.2}\, e^{-2\pi i \, \mathrm{OPD}/\lambda + i\,\Phi^\mathrm{off}(\vec{s},\lambda)} d\lambda \,\,,
\end{equation}
and the flux ratios $f_k$, $f_l$ in Eq.~\ref{eqA1} are multiplied with the ratio of $A^\mathrm{off}$ for the two sources.

 \begin{figure}[h]
\centering
\includegraphics[width=9cm]{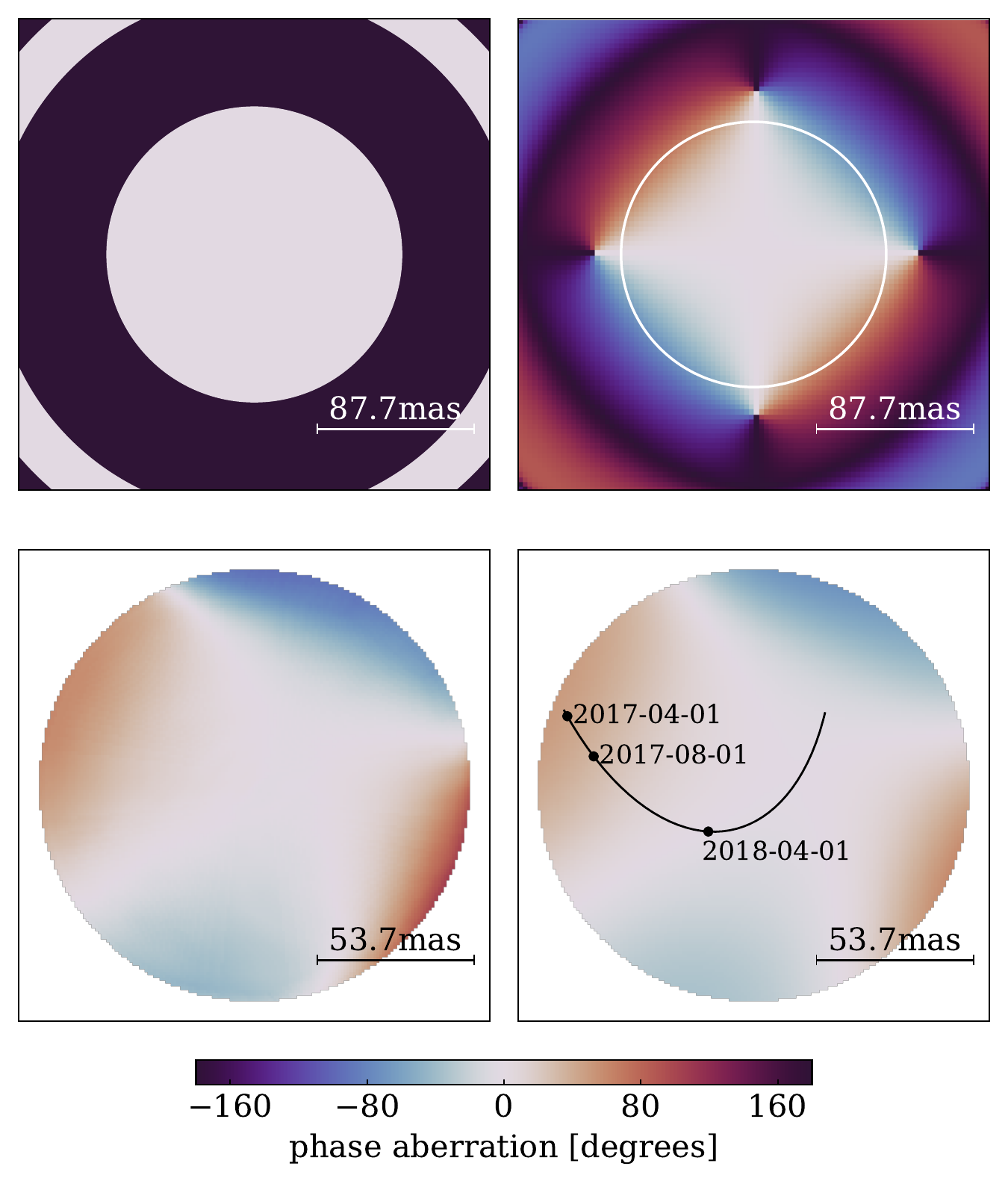}
\caption{Two-dimensional phase maps used for correcting the effects of static aberrations in the binary fitting. The top row compares simulations of a perfect Airy pattern (left) to a static astigmatism with $180\,$nm RMS over the full pupil. The white circle in the top right panel shows the extent of the measured phase maps. One example for such a map is shown below, before (left) and after (right) applying a Gaussian kernel accounting for atmospheric smoothing. The black line in the lower right panel indicates the trace where S2 was located, as predicted by the orbit.
}
\label{fig:figA4}
\end{figure}

 \begin{figure}[h]
\centering
\includegraphics[width=9cm]{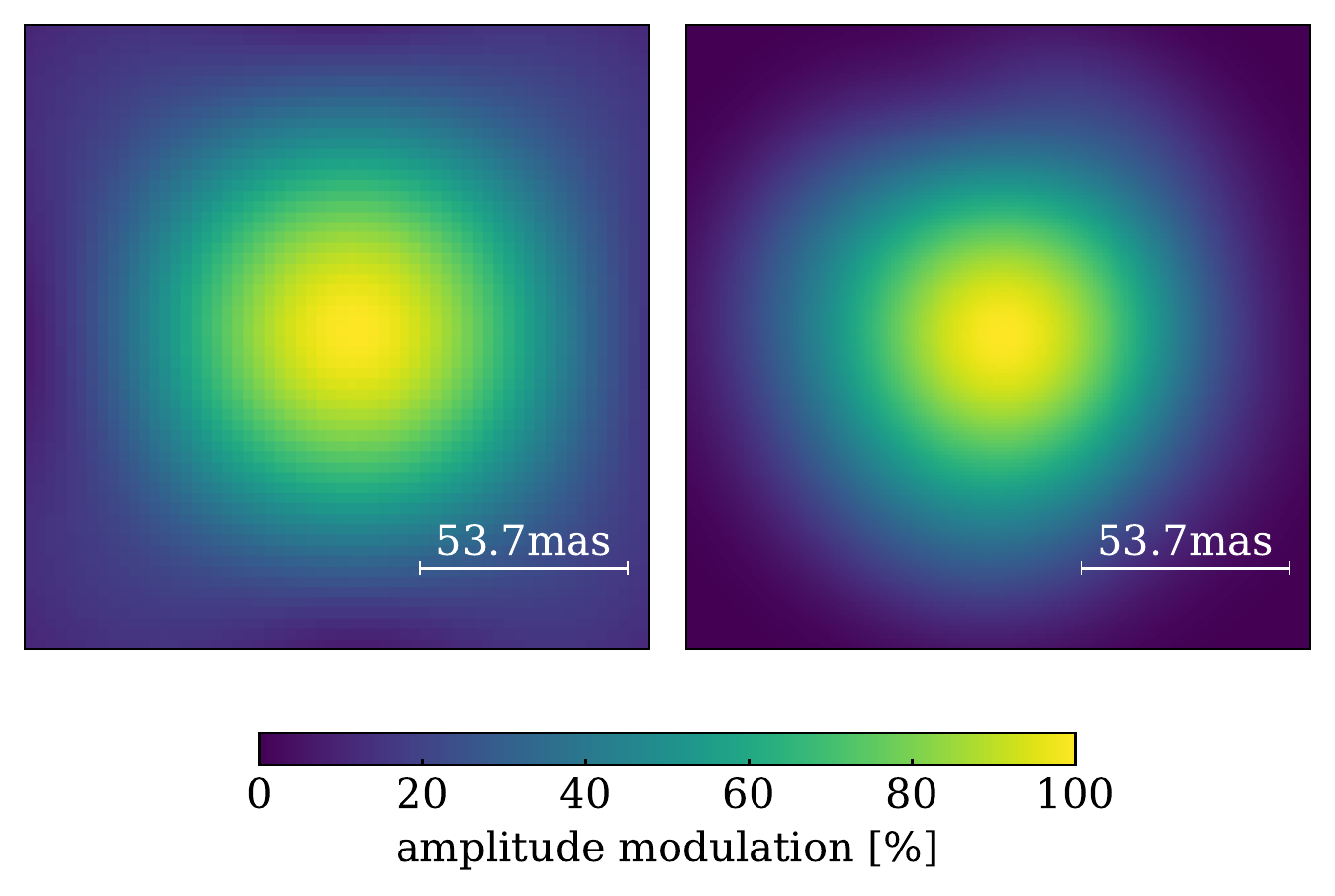}
\caption{Comparison of the theoretical amplitude map smoothed with atmospheric AO residuals (left) and the measured map smoothed with a Gaussian kernel to match the on-sky measured width (right). The FWHM for the left panel is $87\,$mas, and for the right panel it is $88\,$mas.}
\label{fig:figA5}
\end{figure}

The refined binary fitting therefore requires maps of the amplitude and phase distortion as additional input. We obtained these from dedicated calibration runs, using the GRAVITY calibration unit, which simulates the light of an unresolved source. The offset between this source and the fibre can be controlled, and we scanned the FOV in order to measure the relative changes in phase and amplitude across (see Figure~\ref{fig:figA4} for an example).

In contrast to an astronomical observation, our calibration data are not affected by the smoothing effects of the AO residuals. We account for the atmospheric smoothing by applying a Gaussian kernel to the phase maps. The typical tip-tilt jitter for observations of the GC has an rms per-axis of $\approx 15\,$mas \citep{2019A&A...625A..48P}, and higher order aberrations also contribute. We determined the amount of atmospheric smoothing by comparing the amplitude maps with the actual on-sky profiles, and verified in a simulation that the effects of a static astigmatism plus atmospheric broadening match the observed widths (Figure~\ref{fig:figA5}). 
The uncertainty in the atmospheric smoothing yields an additional systematic error for the astrometry that we assessed by using different smoothing kernels, which result in an FWHM of the amplitude map between $88$ and $96\,$mas.

The effect of the static aberration does not average out, because the orientation of the field inside GRAVITY is always the same for our observations. Moreover, the projected baselines are not drawn from full tracks in the uv-plane, but we rather have a typical observing geometry. We therefore expect a bias. In comparison to binary fits neglecting static aberrations, we find that the position of S2 is indeed offset systematically throughout 2017 by approximately $0.44 \times (t - 2018) - 0.10\,$mas in RA and $-0.86 \times (t - 2018) + 0.28\,$mas in Dec. As expected, the offset decreases as S2 moves closer to Sgr~A*.
Finally, we note that our result for $f_\mathrm{SP}$ does not depend in a significant way on this correction.

\section{Theoretical expectations for the precession of S2}
\label{theo}
The $12.1'$ precession angle predicted by GR corresponds to a spatial shift between the GR and the Kepler orbit of $0.78\,$mas at apocentre, mostly in RA because of the current orientation of the orbit. To detect this shift with $5\sigma$ significance requires a positional measurement precision of $100\,\mu$as or less. We have more than 100 NACO measurements of the orbit, each with a statistical precision of $400\,\mu$as. If we did not have systematics (offset and drift of the infrared to mass-radio references frames) it should therefore (have) be(en) possible to detect the SP with NACO or the Keck NIRC imager alone. While the motion on the sky of S2 could be detected with NACO over periods of months, the GRAVITY observations detect the star's motion over $0.5-2\,$days.  

\begin{figure}[h]
\centering
\includegraphics[width=6.8cm]{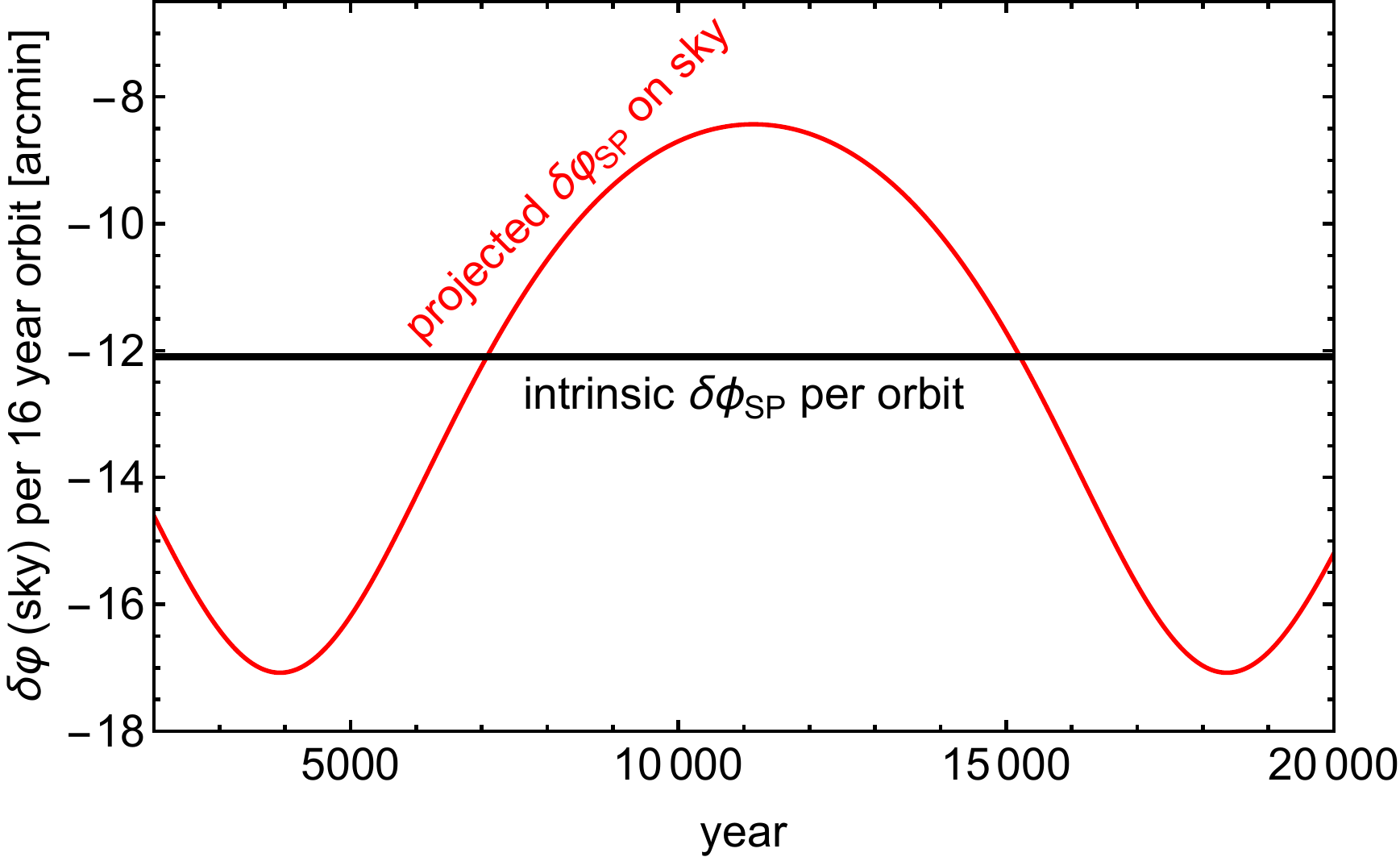}
\caption{Advance of sky-projected apocentre angle $\delta \varphi$, per orbital period, as a function of year.}
\label{fig:figB1}
\end{figure}

The precession angle projected on the sky depends on the geometric angles of the orbit and therefore on time. In the plane of the orbit, the precession advances the angle $\delta \phi$ by $12.1'$ per orbital period of 16.046 yr. The precession projected on the sky $\delta \varphi$ varies from $-17'$ to $-8.4'$ through each half SP period of $P_\mathrm{SP} = 28,710\,$yr (Figure~\ref{fig:figB1}).

\begin{figure*}[h]
\centering
\includegraphics[width=13cm]{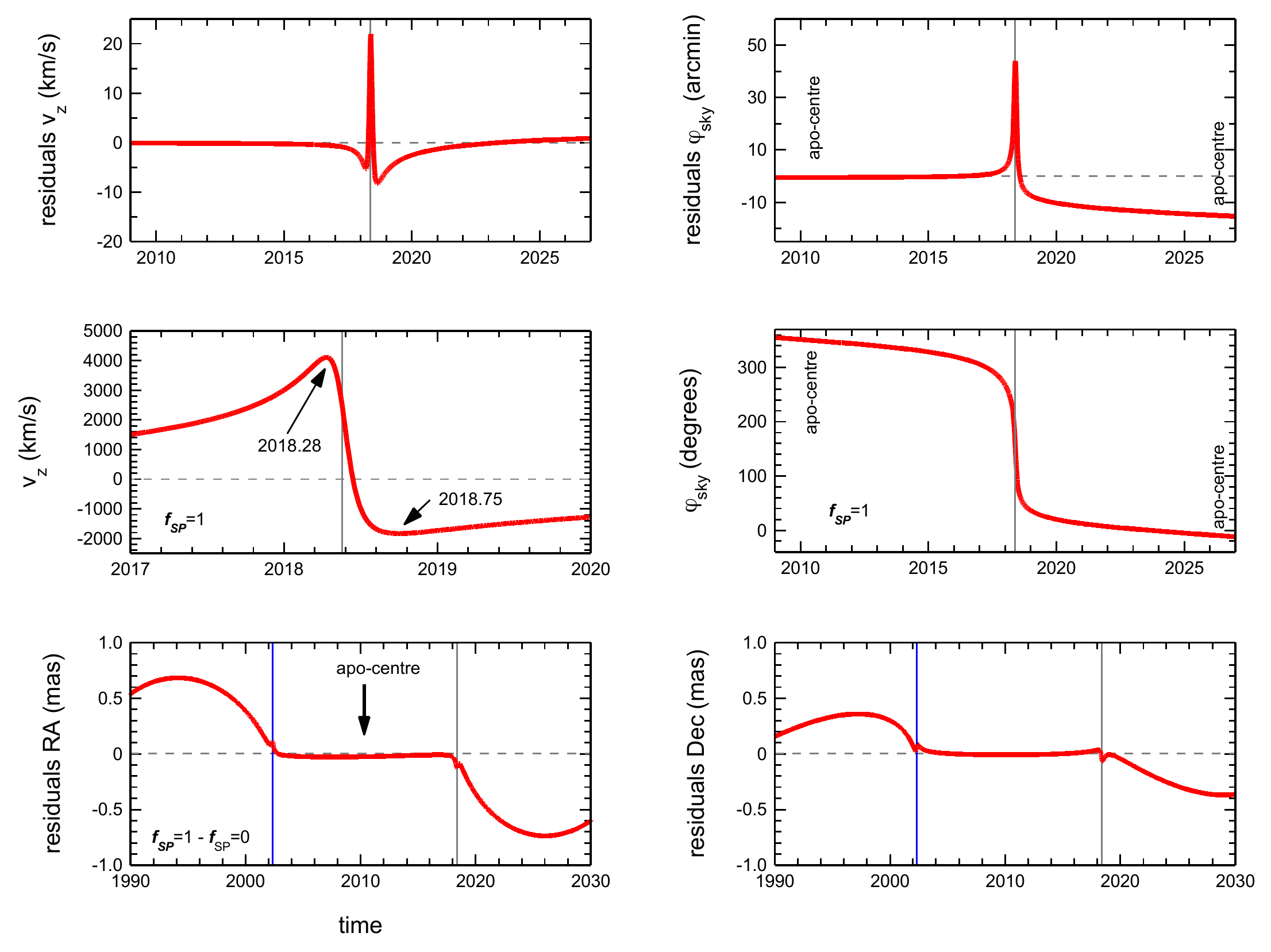}
\caption{Theoretical expectations for the effect of the Schwarzschild precession on the orbit of the star S2. Here we took the best-fit parameters of the S2 orbit, and computed two model orbits, one for $f_\mathrm{SP} = 0$ (Newton, plus R{\o}mer effect, plus SRT, plus RS), and one for $f_\mathrm{SP} = 1$ (equation~\ref{eqB1}). The grey (2018.38) and blue (2002.34) vertical lines are the pericentre times. We arbitrarily set the precession angle of the SP orbit to 0 during apocentre 2010.35. The top panels denote the residuals of $\delta vz$ (left) and $\delta \varphi$ (right) 
between the $f_\mathrm{SP} = 1$ and $f_\mathrm{SP} = 0$ orbits. The bottom panels show the same for $\delta$RA (left) and $\delta$Dec (right). The middle panels present $vz$ (left) and $\varphi$ (right) as a function of time. Here, $\varphi$ is the position angle of the star on the sky, $\varphi = \arctan(\mathrm{RA}/\mathrm{Dec})$, running from $359^\circ$ when the star is straight north, or north-west of centre, to $180^\circ$ when it is straight south, to $>0^\circ$ when it is just north-north east of centre. The most fundamental aspect of the precession is seen in the top right panel as a change in $\delta \varphi$ by $\approx14'$ between two apocentres. Because the precession strongly depends on radius, the precession is very fast around pericentre (2018.38) in a highly elliptical orbit, so that within $\approx1\,$year of pericentre $\approx 75$\% of the precession has occurred. To first order, the precession leads to a change in time when the star is found at a given angle $\varphi$ on the sky, relative to the non-precessing orbit. Because the functional form of $\varphi(t)$ is close to a step function, the differencing $\delta \varphi(t) = \varphi_\mathrm{SP=1}(t) -\varphi_\mathrm{SP=0}(t)$ is close to a differentiation $d\varphi/dt$, which thus results in a sharp $\delta$-function in the residuals $\delta \varphi(t)$ near pericentre. In velocity space a similar effect occurs in the residuals as well, although $vz(t)$ is not as symmetric in $t$ relative to $t_\mathrm{peri}$. Finally in $\delta$RA and $\delta$Dec (bottom panels), the effect of the precession results in a `kink' in the orbit coordinate time slope. Because of the variations in the foreshortening of the RA and Dec coordinates of the apocentre values of the $\delta$RA, $\delta$Dec and $\delta \varphi$ the SP = 1 vs. SP = 0 curves vary over time (Figure~\ref{fig:figB1}). The projected precession on the sky between the apocentres 2010.35 and 2026.5 is $\approx 14'$. }
\label{fig:figB2}
\end{figure*}

Figure~\ref{fig:figB2} illustrates the effects the SP is expected to have on the measured parameters of the S2 orbit. Because of the strong dependence of $\delta \phi$ on radius, much of the $12.1'$ precession occurs within $\pm 1\,$year of pericentre. In RA/Dec space, the precession is seen as a `kink' in the time change of the post-pericentre versus pre-pericentre residuals.  
Very near pericentre passage, the precession acts to first order as a time-shift between the precessing and the equivalent $f_\mathrm{SP}=0$  orbit (see also Fig.~\ref{fig:figE2} top left panel). In the residuals $\delta$RA, $\delta$Dec, $\delta$vz, and $\delta \varphi$ between the data and the $f_\mathrm{SP}=0$, short-term excursions of about a few times $\beta^2$ appear in all these observables as a result.

\section{Parametrisation of the Schwarzschild precession}
\label{sec:PPN}

We uses the post-Newtonian limit for the equation of motion presented in \cite{2008ApJ...674L..25W}, Eq.~(1) therein. We parametrised the effect of the Schwarzschild metric (i.e. the prograde precession) by introducing an ad hoc factor $f_\mathrm{SP}$ in front of the terms arising from the Schwarzschild metric. We set the terms due to spin $J$ and quadrupole moment $Q_2$ to 0. This results in
\begin{equation}
\vec{a} = -\frac{G M}{r^3} \vec{r} 
+ f_\mathrm{SP} \frac{G M}{c^2 r^2}\left[\left(4 \frac{G M}{r} - v^2\right)\frac{\vec{r}}{r} +4 \dot{r}\vec{v}\right] + O[J] + O[Q_2] \,\, .
\label{eqB1}
\end{equation}
In the (first-order) parameterized post-Newtonian (PPN) expansion of GR, the second term becomes
\begin{equation}
\frac{G M}{c^2 r^2}\left[\left(2 (\gamma+\beta) \frac{G M}{r} - \gamma v^2\right)\frac{\vec{r}}{r} +2(1+\gamma) \dot{r}\vec{v}\right] \,\,.
\end{equation}
In GR, $\beta_\mathrm{GR} = \gamma_\mathrm{GR} = 1$. Two PPN parameters $(\beta, \gamma)$ are needed to describe the equation of motion, and for no choice of $\beta$ and $\gamma$ can we recover the Newtonian solution. The PPN formalism is therefore less well suited for our experiment than using $f_\mathrm{SP}$.

The net effect of the precession is
\begin{equation}
\Delta \phi_\mathrm{per\,orbit} = 3 f_\mathrm{SP} \frac{\pi R_S}{a(1-e^2)}\,\,,
\end{equation}
for our parametrisation, and to
\begin{equation}
\Delta \phi_\mathrm{per\,orbit} =  (2+2\gamma-\beta)  \frac{\pi R_S}{a(1-e^2)}\,\,,
\end{equation}
in the PPN formulation of GR \citep{2014LRR....17....4W}. Yet it is imprecise to identify the factor $3 f_\mathrm{SP}$ with $ (2+2\gamma-\beta)$. Our parameter $f_\mathrm{SP}$  characterises how relativistic the model is, as is probably easiest seen by the fact that its effect corresponds to changing the value of the speed of light in the equations, with the limit
$c \rightarrow \infty \,\, \mathrm{for} \,\, f_\mathrm{SP} \rightarrow 0$. 

\begin{figure}[h]
\centering
\includegraphics[width=7cm]{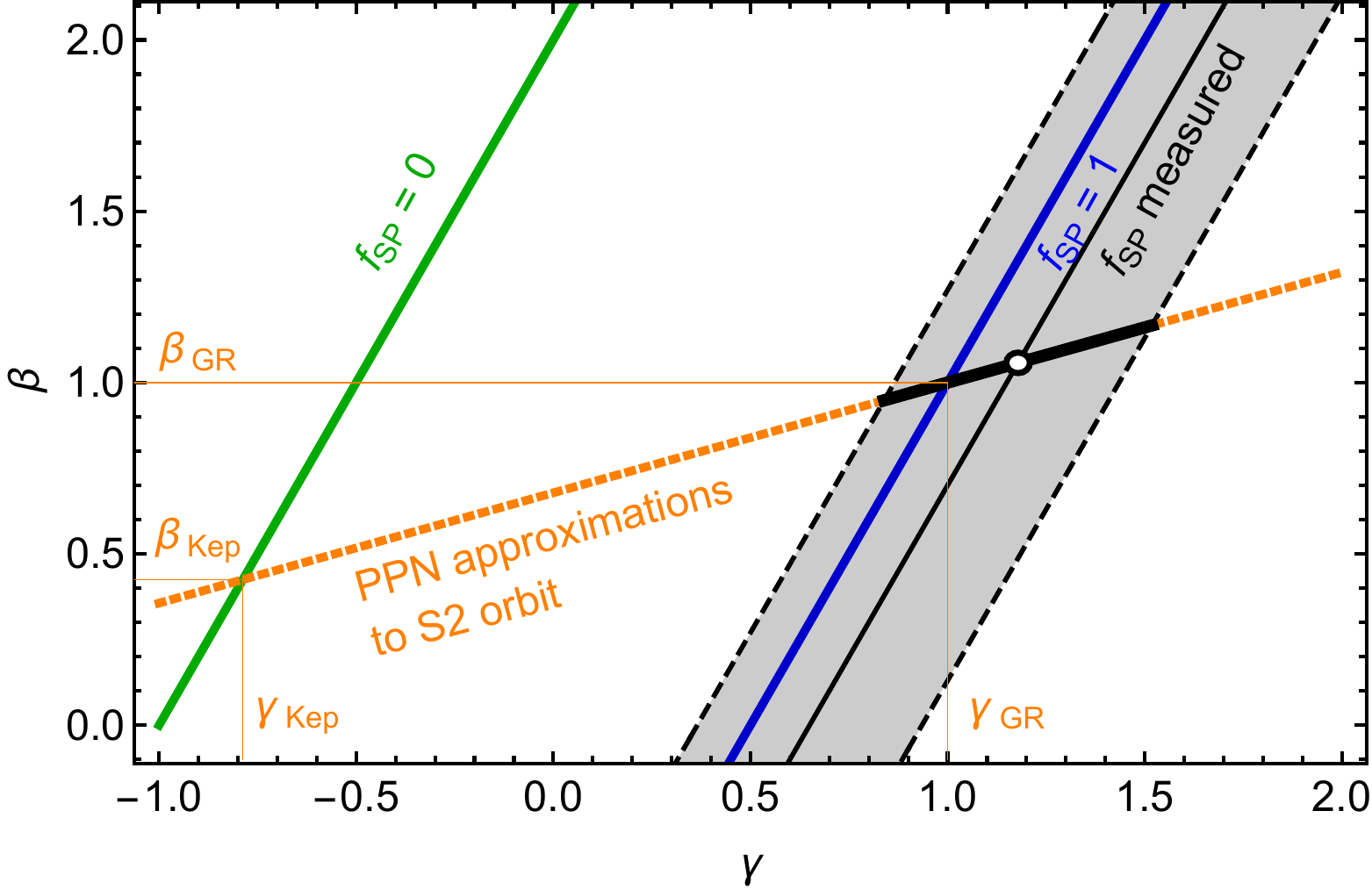}
\caption{Interpretation of our measurement in the plane of the PPN parameters $\beta$ and $\gamma$. Our value for $f_\mathrm{SP}$ and its uncertainty are represented by the black line and grey band. The GR value $f_\mathrm{SP}=1$ is the blue line, and the Newtonian value  $f_\mathrm{SP}=0$ is the green line. The best approximations to the orbits by PPN parameters are shown by the orange dotted line. Assuming GR is a PPN theory, our measurement corresponds to the white circle at the intersection point and the uncertainties are the adjacent black thick lines.}
\label{fig:figC1}
\end{figure}

In the PPN formulation of GR, all orbits with $\beta = 2(1 + \gamma)$ have zero net precession per revolution, and all orbits with $\beta = 2\gamma - 1$ have the same amount of pericentre advance as GR.
Because of the high eccentricity of the S2 orbit, the precession leads to an almost instantaneous change of the orbit orientation $\omega$ in its plane when the star passes pericentre. Our result therefore essentially compares the orbit orientations post- and pre-pericentre 2018. In this limit, we can indeed state that we have measured $(2+2\gamma-\beta)/3 = 1.10 \pm 0.19$. Figure~\ref{fig:figC1} illustrates our constraint in the plane spanned by $\beta$ and $\gamma$. Because there is no exact representation of the Keplerian orbit in the PPN formalism, we instead seek the PPN orbit that most closely resembles the Keplerian orbit. This depends on the eccentricity, and for S2, we find  $\gamma_\mathrm{Kep} = -0.78762$ and  $\beta_\mathrm{Kep} = 0.42476$. Changing  $f_\mathrm{SP}$  corresponds  to moving along a line from $(\gamma_\mathrm{Kep}, \beta_\mathrm{Kep})$ to $(\gamma_\mathrm{GR}, \beta_\mathrm{GR})$.  With this, we find $\beta = 1.05 \pm 0.11$ and $\gamma = 1.18 \pm 0.34$, and the two are fully correlated.

\section{Astrophysical implications}
\label{sec:discussion}

\begin{figure}[h]
\centering
\includegraphics[width=8.4cm]{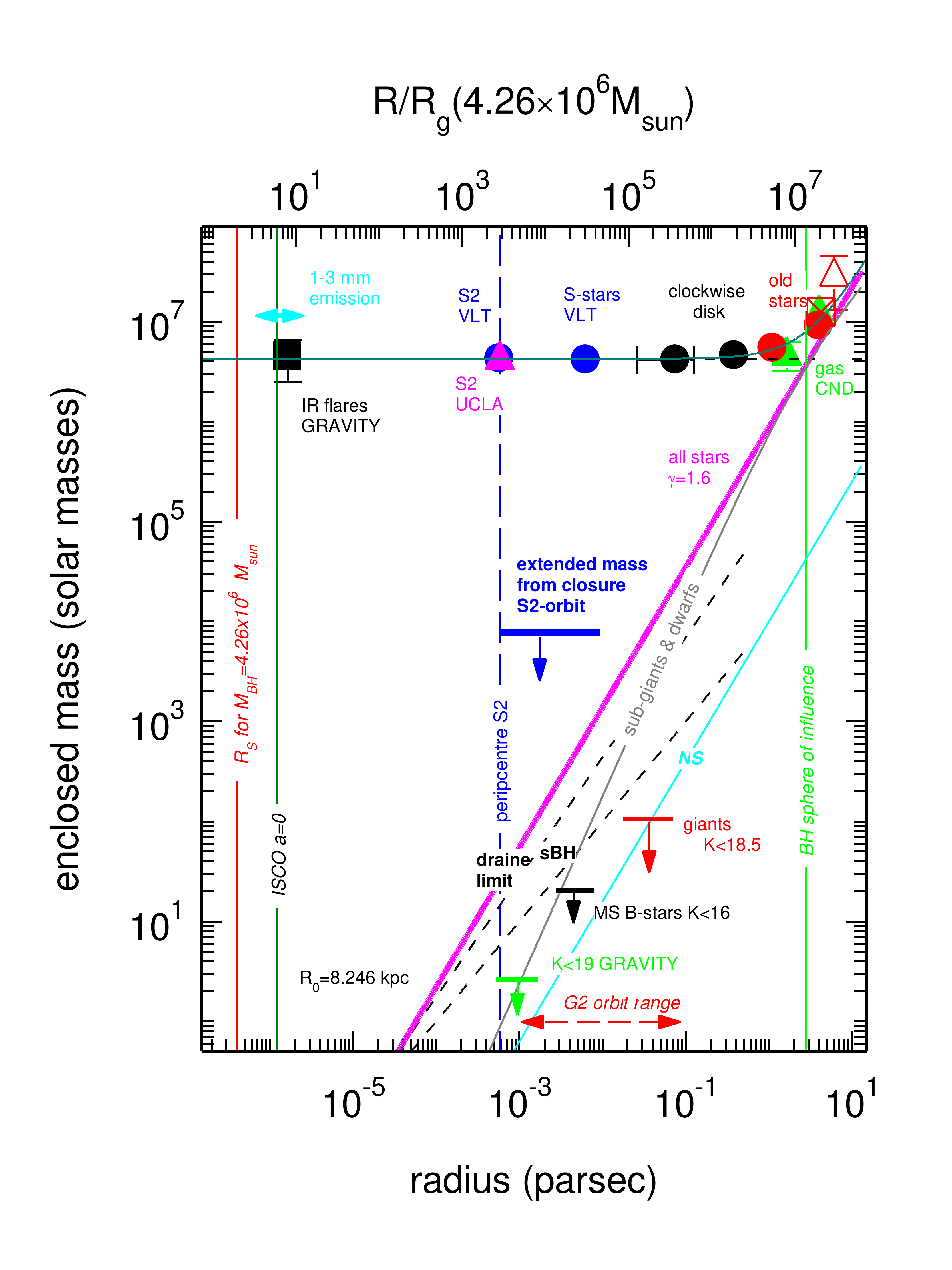}
\caption{Constraints on the enclosed mass in the central $10\,$pc of the Galaxy. The blue crossed circle, the pink triangle, and the black crossed rectangles are estimates of the enclosed mass within the S2 orbit, other S-stars and the massive star discs \citep{2006ApJ...643.1011P, 2009ApJ...697.1741B, 2014ApJ...783..131Y}. The red filled circles, the red crossed rectangle, and red open triangles denote mass measurements from late-type stars. Green triangles are mass estimates from rotating gas in the circum-nuclear disc (see \citealt{2010RvMP...82.3121G} for details). The filled black rectangle comes from the clockwise loop-motions of synchrotron near-infrared flares \citep{2018A&A...618L..10G}. The cyan double arrow denotes current VLBI estimates of the $3\,$mm size of Sgr~A* \citep{2019ApJ...871...30I}. The continuous magenta line shows the total mass from all stars and stellar remnants \citep{2017ARA&A..55...17A}. The grey line marks the distribution of $K<18.5$ sub-giants and dwarfs from \cite{2018A&A...609A..27S}. The black dashed lines and the cyan line indicate the distribution of stellar black holes and neutron stars from theoretical simulations of \cite{2017ARA&A..55...17A} and \cite{2018A&A...609A..28B}, which span a range of roughly a factor 5. Red, black and green upper limits denote upper limits on giants, main-sequence B stars and $K<19$ GRAVITY sources. The Schwarzschild radius of a $4.26\times 10^6 M_\odot$ black hole and the innermost stable circular orbit radius for a non-spinning black hole are given by red vertical lines. The pericentre radius of S2 is the dashed vertical blue line and the sphere of influence of the black hole is given by the vertical green line. The blue horizontal line denotes the $2\sigma$ upper limit of any extended mass around Sgr~A* obtained from the lack of retrograde precession in the S2 orbit (see text).}
\label{fig:figD1}
\end{figure}

\begin{figure}[h]
\centering
\includegraphics[width=8cm]{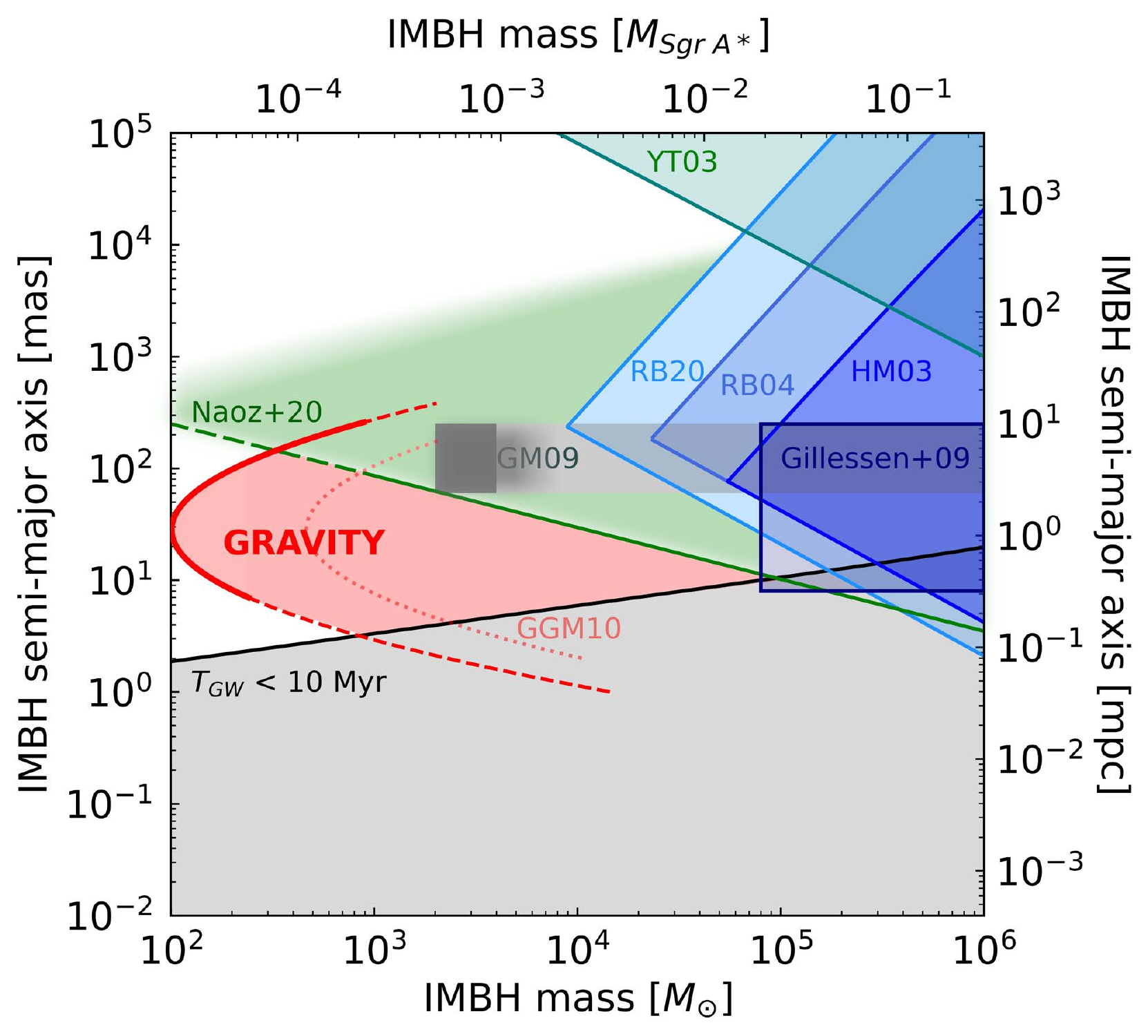}
\caption{Limits on a second, intermediate mass black hole (IMBH), as a function of its mass and separation from Sgr~A*. The shaded area is excluded observationally. Adapted from \cite{2009ApJ...705..361G}. The blue shaded regions are due to the lack of observed motion of Sgr~A*  at radio wavelengths (\citeauthor{2003ApJ...593L..77H} \citeyear{2003ApJ...593L..77H}, HM03). 
The data from  \citeauthor{2004ApJ...616..872R} (\citeyear{2004ApJ...616..872R}, RB04) and  \citeauthor{2020arXiv200104386R} (\citeyear{2020arXiv200104386R}, RB20) improve these limits. 
The upper bound results from the limit on the 3D velocity
$v_\mathrm{3D} \lesssim 8\,$km/s (RB04) and $\lesssim 3\,$km/s (RB20). 
The lower bound results from the absence of short-period fluctuations in the position, with limits of $1\,$mas 
and $0.5\,$mas in RB04 and RB20.
\cite{2003ApJ...599.1129Y} set a limit from the ejection rates of hypervelocity stars if an IMBH were present in the GC (YT03). The bottom area is excluded because the gravitational wave inspiral time scale $T_\mathrm{GW}$ would be  $< 10\,$Myr. \cite{2020ApJ...888L...8N} exclude the green shaded area by demanding that the S2 orbit be stable, taking into account resonant effects from the IMBH. Beyond the ranges given in the original work, the constraints get weaker (fading color). The area labeled GM09 is excluded by \cite{2009ApJ...705..361G}, who calculated the effect of an IMBH on the distribution of stellar orbits. Their original box extends to higher masses (shaded area to right). A first constraint from the orbit data of S2 was given in \cite{2009ApJ...692.1075G}. Extending the orbital coverage in a simulation to the 2018 pericentre passage \cite{2010MNRAS.409.1146G} concluded that from a lack of extra residuals one should be able to exclude the area right of the dotted line  (GGM10). GRAVITY improved the accuracy compared to these simulations by a factor $4.6$, which moves the limit further to lower masses (red-shaded region). All but a $10^2-10^3 M_\odot$ IMBH inside or just outside of S2's orbit is now excluded by the various measurements. This also excludes the configurations \cite{2009ApJ...693L..35M} found to be efficient in randomizing the S-star orbits.}
\label{fig:figD2}
\end{figure}

{\bf Distributed mass component inside the orbit of S2}: An extended mass component would create retrograde Newtonian precession. Our data strongly constrain such a component. For simplicity we use spherically symmetric distributions of the extended mass. Using a \cite{1911MNRAS..71..460P} profile with a scale parameter of 0.3 arcseconds \citep{2005AN....326...83M} and fitting for the normalisation of that mass component assuming $f_\mathrm{SP} = 1$ shows that $(0.00 \pm 0.10)$\% of the central mass could be in such an extended configuration. Changing the radius parameter to 0.2 or 0.4 arcseconds yields $(-0.02 \pm 0.09)$\% or $(0.01 \pm 0.11)$\%. Using instead a power-law profile with logarithmic slope between $-1.4$ and $-2$ results in a mass estimate of $(-0.03 \pm 0.07)$\%.
Overall, we estimate that for typical density profiles the extended mass component cannot exceed $0.1$\%, or $\approx4000 M_\odot$ ($1\sigma $ limits). For comparison, modelling of the star cluster suggests that the total stellar content within the apocentre of S2 is $<1000 M_\odot$, and the mass of stellar black holes within that radius is $80 - 340 M_\odot$ (Figure~\ref{fig:figD1}, cf. \citealt{2010RvMP...82.3121G, 2017ARA&A..55...17A, 2018A&A...609A..28B}). We conclude that the expected stellar content within the S2 orbit is too small to significantly affect the SP.

\cite{2010PhRvD..81f2002M} investigated for which configurations the Newtonian precession due to an extended mass component in the form of individual stellar mass objects exceeds the effects of spin and quadrupole moment of the MBH. They addressed a range of masses between $1$ and $10^3\,M_\odot$ in the central milli-parsec. The above limits translate into a limit of $\approx 200\,M_\odot$ in that radial range. Figure~1 of \cite{2010PhRvD..81f2002M} shows that for S2 itself, our limit on the extended mass would lead to perturbations almost on par with the expected spin effects for a maximally spinning MBH, giving some hope that the spin of Sgr~A* can eventually be detected from S2 despite its large orbital radius. \cite{2017ApJ...834..198Z} cautioned, however, that already the Newtonian perturbation from S55/S0-102 \citep{2012Sci...338...84M, 2017ApJ...837...30G} might hide the spin's signature.
For stars on shorter period orbits or with higher eccentricities, detecting the higher order effects of the metric is easier; and stellar perturbations have a different observational signature than the effect of the metric.

{\bf A second massive object in the GC}: The presence of an intermediate mass black hole (IMBH) orbiting Sgr~A* inside the orbit of S2 is constrained by our measurements. \cite{2010MNRAS.409.1146G} explored a grid of three-body simulations with an IMBH of mass 400 to $4000\,M_\odot$ on orbits similar in size to the S2 with a range of angles and eccentricities relative to the S2 orbit. By inspecting the 
astrometric and spectroscopic residuals from the three-body system in comparison to the assumed astrometric error, they concluded that the 2018 pericentre passage of S2 would exclude a large part of the parameter space.
The additional data since 2010 and the much more accurate astrometry from GRAVITY now exclude any IMBH greater than about $1000\, M_\odot$ in the central arcsecond, and allow IMBHs in the mass range of $100-1000 \,M_\odot$ only in a small region inside or just outside of the orbit of S2 (Figure~\ref{fig:figD2}). In the radial regime of the stellar discs ($1'' - 10''$) an IMBH of up to $10^4 M_\odot$ is still allowed.

{\bf The distance to the GC}: Our data set continues to constrain $R_0$ ever better. Setting $f_\mathrm{SP} = f_\mathrm{RS} = 1$ fixed during the fit, we obtain our best estimate for $R_0 = 8248.6 \pm 8.8\,$pc, where the error is the statistical error alone. This is 25\% more precise than our result in \cite{2019A&A...625L..10G}, but the values differ by $\approx 2 \sigma$ when we take the systematic error from \cite{2019A&A...625L..10G} into account. 
We now conservatively adopt, with the improved data set, a systematic error of $45\,$pc, which is twice as large as before. This better reflects  the variations between \cite{2018A&A...615L..15G}, \cite{2019A&A...625L..10G} and this work. Our current best estimate is therefore $R_0 = 8249 \pm 9|_\mathrm{stat.} \pm 45|_\mathrm{sys.}\,$pc. 
Because of the strong correlation between the best-fit MBH mass $M_\bullet$ and $R_0$ (Figure~\ref{fig:figE2}), the increase in $R_0$ is reflected in $M_\bullet$.

\begin{figure}[h!]
\centering
\includegraphics[width=8.05cm]{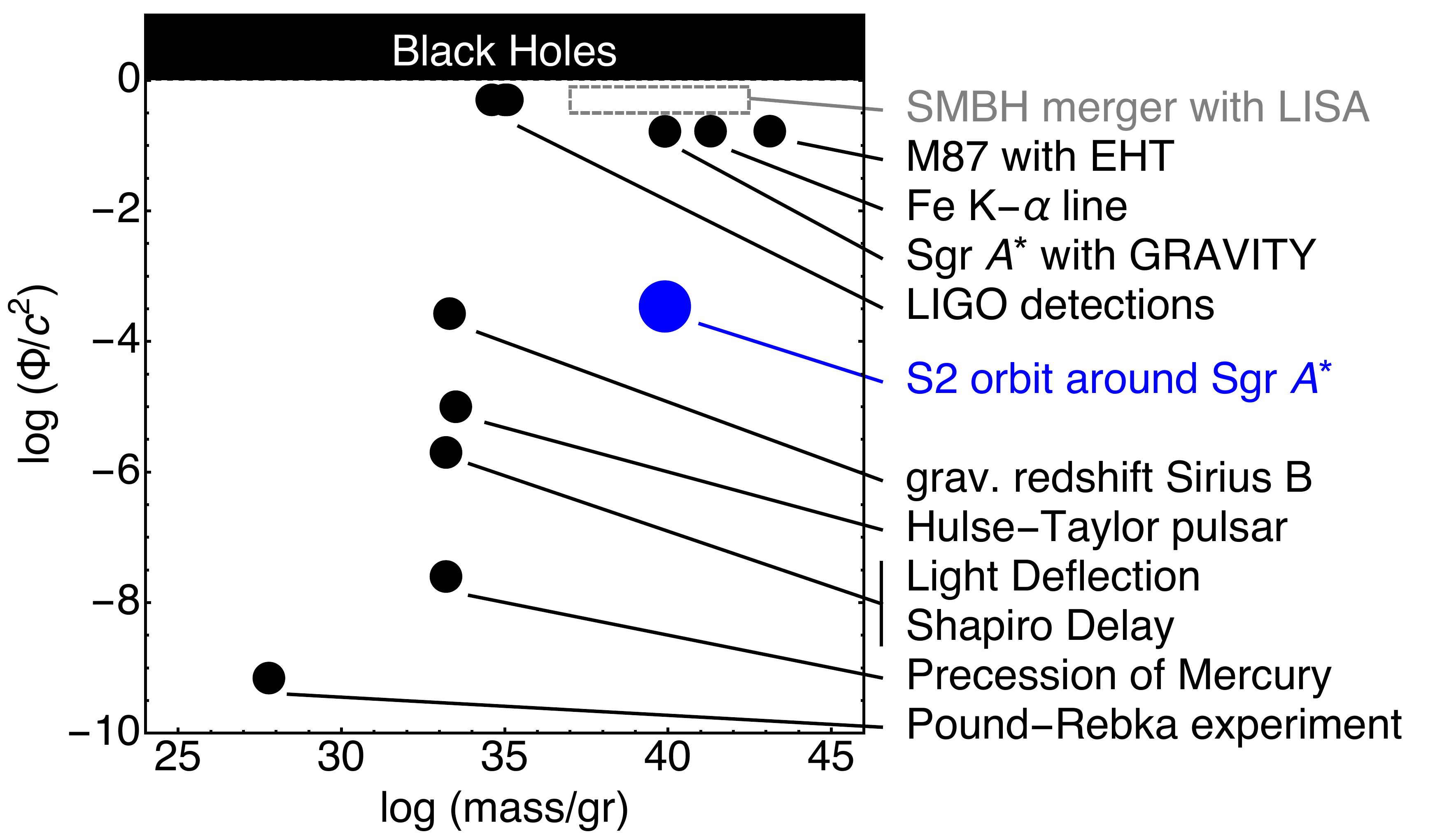}
\caption{Comparison of tests of GR in the plane of mass and potential, adapted from \cite{2004AIPC..714...29P}. Black: Terrestrial laboratories, Mercury's precession, light deflection, and Shapiro delay in the Solar System, the Hulse-Taylor pulsar, the LIGO detections, the relativistic K-$\alpha$ lines, and the M87 EHT observation. LISA signals will probe the grey rectangular region. This work, using S2, is marked in blue.}
\label{fig:figD3}
\end{figure}

{\bf Constraints on PPN parameters}: In Appendix~\ref{sec:PPN} we derived our constraints on the PPN parameters $\beta = 1.05 \pm 0.11$ and $\gamma = 1.18 \pm 0.34$. These are consistent with GR, but not competitive with the results obtained in the Solar System from spacecraft measurements. The deviation from the GR value of $\gamma = 1$ is best constrained through the Shapiro delay from the Cassini spacecraft to better than $2\times 10^{-5}$, while VLBI measurements of the light deflection using the quasars 3C273 and 3C279 yield a factor 10 weaker constraints on $\gamma$ \citep{2014LRR....17....4W}. Assuming the value of $\gamma$ from Cassini, the SP of Mercury's orbit from the Messenger spacecraft yields a constraint on $\beta$ of $8\times 10^{-5}$. While our constraints are weaker, they probe a completely different regime in mass (by a factor $4\times 10^6$) and potential strength (by a factor $10^2$ to $10^4$) than the Solar System tests (Figure~\ref{fig:figD3}).

{\bf Beyond the standard model}: We also derived limits on a Yukawa-like potential in the GC \citep{2017PhRvL.118u1101H}. Our limits show the same sensitivity to the length-scale parameter $\lambda$ as in \cite{2017PhRvL.118u1101H}, but are a factor 20 more constraining in terms of the interaction strength $\alpha$. At our most sensitive $\lambda = 180\,$AU, we constrain $|\alpha|< 8.8\times 10^{-4}$ (95\% confidence level).

Our data also constrain the possible parameters for an assumed dark-matter spike in the GC. \cite{2018A&A...619A..46L} showed that for a Navarro-Frenk-White profile \citep{1996ApJ...462..563N} with slope $\gamma=-1$ plus a spike with a power-law profile with slope $\gamma_\mathrm{sp} = -7/3$, the data of \cite{2017ApJ...837...30G} constrain the spike radius to $R_\mathrm{sp} \lesssim 100\,$pc, corresponding to an enclosed mass of $\approx 5\times 10^4\,M_\odot$. Our new data set constrains $R_\mathrm{sp} \lesssim 10\,$pc (corresponding to  $\approx 3\times 10^3\,M_\odot$), which is just below the theoretical prediction from \cite{1999PhRvL..83.1719G}, who took limits from the absence of a neutrino signal from the GC into account.

\section{Details of the fit}
\label{details}
\begin{figure}[!h]
\centering
\includegraphics[width=5cm]{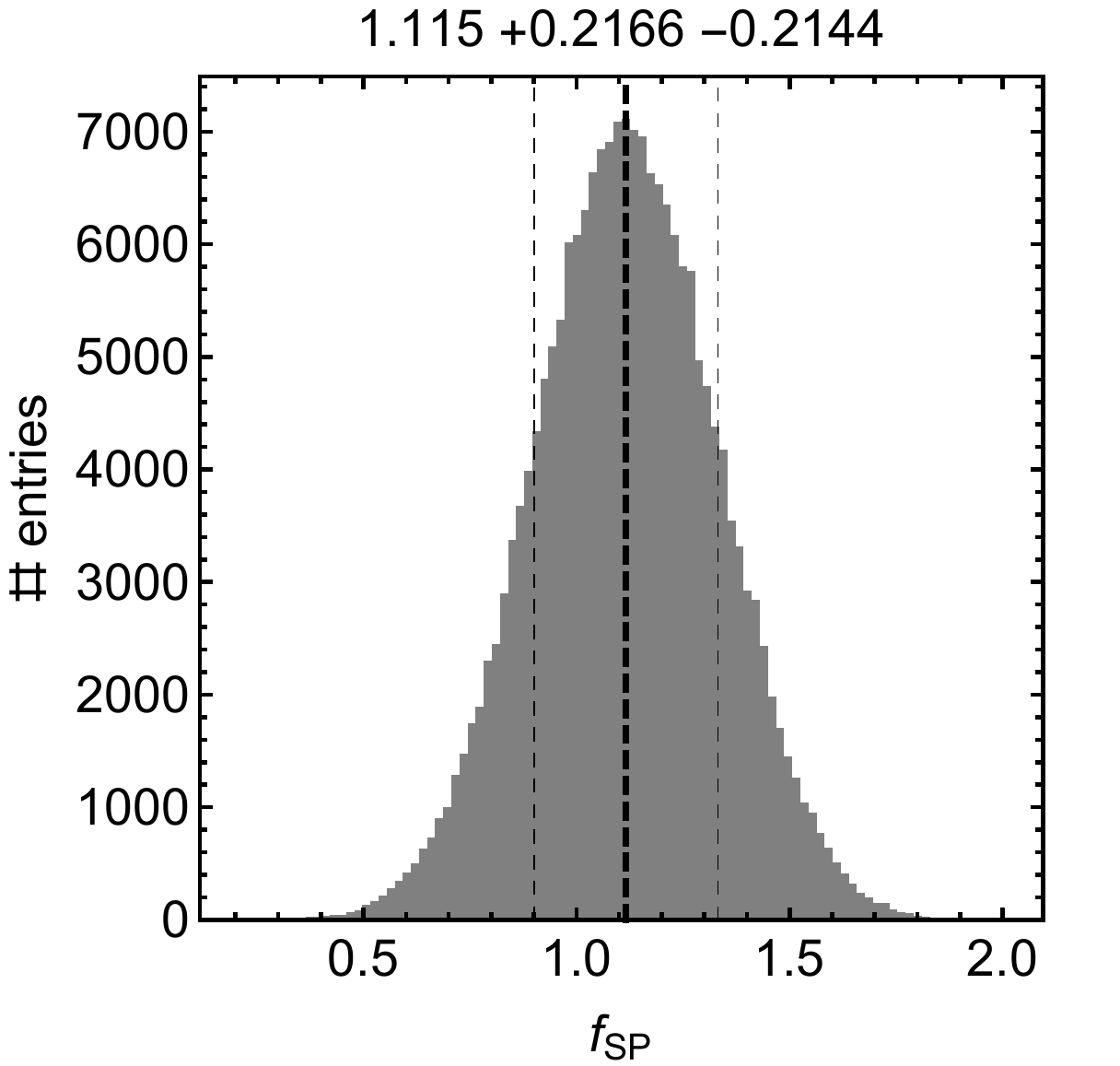}
\caption{Result of the MCMC modelling of our data, showing the posterior distribution of the Schwarzschild parameter $f_\mathrm{SP}$.}
\label{fig:figE1}
\end{figure}

In Table~\ref{tab:t1} we report the best-fitting parameters of our 14-parameter fit, together with the formal fit errors and the $1\sigma$ confidence intervals from the MCMC. The two approaches agree because our fit is well behaved. There is a single minimum for $\chi^2$, and the posterior distribution is close to a 14-dimensional Gaussian (Figure~\ref{fig:figE3}), with significant correlations, however. Figure~\ref{fig:figE1} shows the posterior for $f_\mathrm{SP}$.
 
  \begin{table}
      \caption[]{Best-fit orbit parameters. The orbital parameters are to be interpreted as the osculating orbital parameters. The argument of periapsis $\omega$ and the time of pericentre passage $t_{\rm peri}$ are given for the epoch of last apocentre in 2010.}
      \label{tab:t1}
          \begin{center}
      {\tiny
        \begin{tabular}{lllll}
            \hline
            \noalign{\smallskip}
            Parameter& Value & fit error & MCMC error & Unit \\
            \noalign{\smallskip}
            \hline
            \noalign{\smallskip}
           $f_\mathrm{SP}$ & 1.10 & 0.19 & 0.21 \\
           $f_\mathrm{RS}$ & 1 & fixed & fixed \\
           $M_\bullet$ & 4.261 & 0.012 & 0.012 & $10^6\,M_\odot$ \\
           $R_0$ & 8246.7 & 9.3 & 9.3 & pc \\
           $a$ & 125.058 & 0.041 & 0.044 & mas \\
           $e$ & 0.884649 & 0.000066  & 0.000079  & \\
           $i$ & 134.567 &0.033 & 0.033 & $^\circ$ \\
           $\omega$ & 66.263 & 0.031 & 0.030 &$^\circ$ \\
           $\Omega$ & 228.171&0.031 & 0.031&$^\circ$ \\
           $P$ & 16.0455 & 0.0013 & 0.0013 & yr \\
           $t_{\rm peri}$ & 2018.37900 & 0.00016 & 0.00017& yr \\     
           $x_0$ & -0.90 & 0.14 & 0.15 & mas \\
           $y_0$ & 0.07 & 0.12 & 0.11 & mas \\
           $vx_0$ & 0.080 & 0.010 & 0.010 & mas/yr \\
           $vy_0$ & 0.0341 & 0.0096 & 0.0096 &mas/yr \\
           $vz_0$ & -1.6 & 1.4 & 1.4 &km/s \\
            \noalign{\smallskip}
            \hline
            \noalign{\smallskip} \noalign{\smallskip}
         \end{tabular} 
         }
          \end{center}
      \end{table}

\begin{figure}[!h]
\centering
\includegraphics[width=8.5cm]{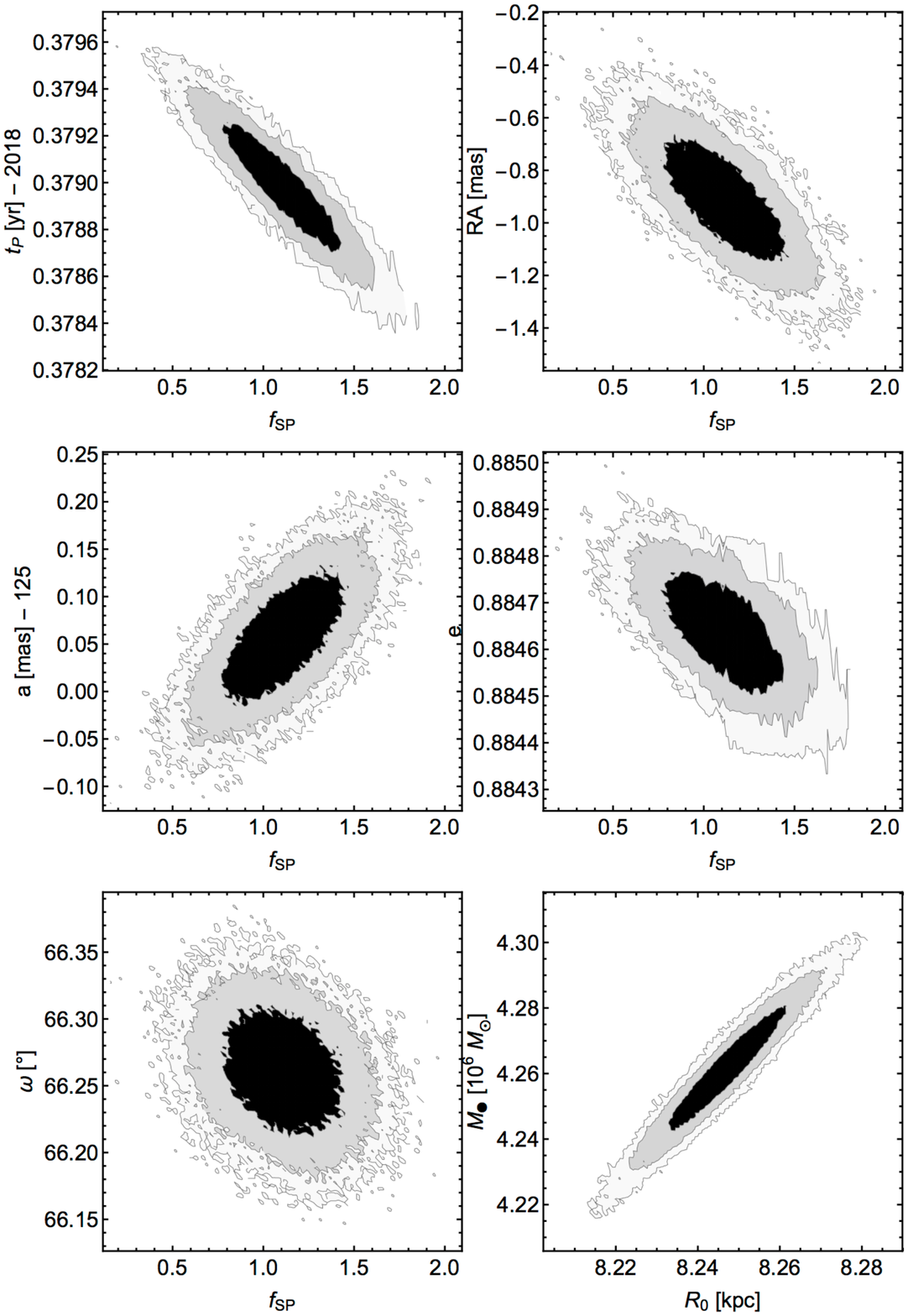}
\caption{Selected parameter correlations from the 14-dimensional posterior distribution as determined from MCMC modelling.}
\label{fig:figE2}
\end{figure}

\begin{figure*}[!h]
\centering
\includegraphics[width=18cm]{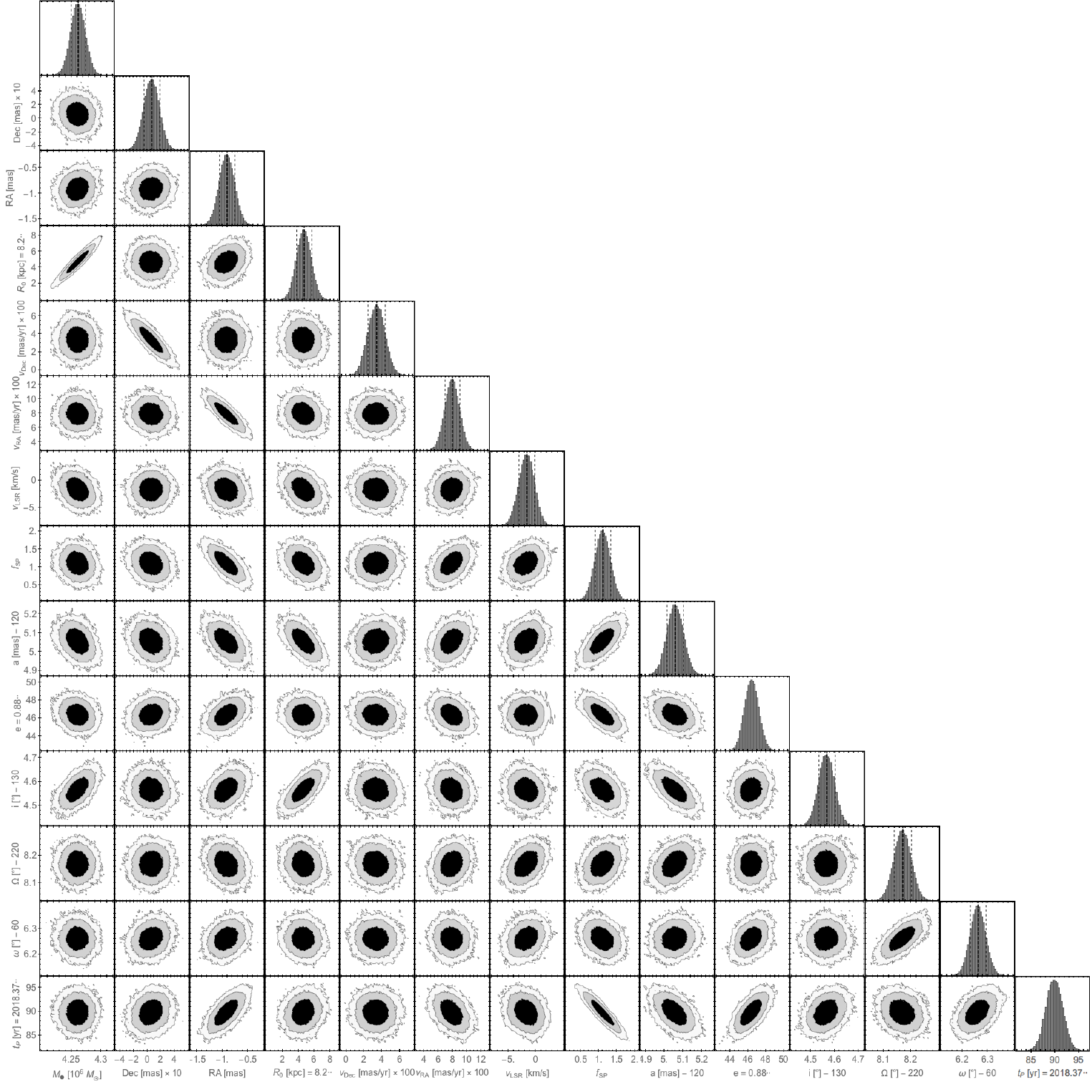}
\caption{Full, 14-dimensional posterior distribution of our orbit fit.}
\label{fig:figE3}
\end{figure*}

In Figure~\ref{fig:figE2} we show selected correlation plots from the posterior distribution, which are worth discussing in the context of $f_\mathrm{SP}$. The strongest correlation for $f_\mathrm{SP}$ is with the pericentre time. This is not surprising, given the discussion in Appendix~\ref{theo}, where we showed that near pericentre the SP acts like a shift in time. The second strongest correlation  for $f_\mathrm{SP}$ is with the RA offset of the coordinate system. This explains why including the NACO flare data helps determining $f_\mathrm{SP}$: The flares essentially measure the offset of the coordinate system. 

The parameter  $f_\mathrm{SP}$ is also weakly correlated with the semi-major axis $a$ and it is anti-correlated with the eccentricity $e$ of the orbit. The former can be understood in the following way: If the orbit were slightly larger on sky, a stronger precession term would be required in order to achieve the same amount of kink (in mas on sky) at pericentre. The latter is understood similarly: A higher eccentricity leads to a narrower orbit figure, and hence less of the precession term would be needed.
Interestingly,  $f_\mathrm{SP}$ is almost uncorrelated with the argument of periapsis $\omega$ (i.e. the angle describing the orientation of the orbital ellipse in its plane), despite that the SP changes exactly that parameter. 

The strongest correlation between any two parameters for our fit is the well known degeneracy between mass $M_\bullet$ and distance $R_0$ \citep{2008ApJ...689.1044G, 2009ApJ...692.1075G,  2016ApJ...830...17B, 2017ApJ...837...30G, 2018A&A...615L..15G, 2019A&A...625L..10G}. The parameter $f_\mathrm{SP}$ is only very weakly correlated with $R_0$.\footnote{{\bf Note added in proof:} After acceptance of this paper we noticed posting \cite{2020arXiv200212598G}. He uses a subset of the data presented here for S2 (measurements taken until 2018), as well as published measurements for the stars S38 and S102, to put constraints on the PPN parameters $\beta$ and $\gamma$. He finds an approximately two-sigma agreement with the value of unity expected in GR. His analysis however fixes the values of the black hole mass and distance, $vx_0$, $vy_0$ and $vz_0$ as well as $x_0$ and $y_0$ in advance, which naturally leads to significantly underestimated error bars. Further, the analysis combines data sets from the VLT and Keck telescopes, without allowing for a coordinate system offset, which however is known to be essential \citep{2009ApJ...707L.114G}.} 

\end{appendix}

\end{document}